\documentclass[reprint,aps,amsmath,amssymb,twoside,prd,showkeys,superscriptaddress]{revtex4-1}
\pdfoutput=1
\usepackage{verbatim}
\usepackage{comment}
\usepackage{graphicx}
\usepackage{graphicx}
\usepackage{epsfig}
\usepackage{bm}
\usepackage{tensor}
\usepackage{amsfonts}
\usepackage{amsmath}
\usepackage{graphicx}
\usepackage{hyperref}
\usepackage[english]{babel} 
\usepackage[T1]{fontenc}
\usepackage[utf8]{inputenc}

\begin{document}
\title{Statistical Nuances in BAO Analysis:  Likelihood Formulations and Non-Gaussianities}

\author{Denitsa Staicova}
\email{dstaicova@inrne.bas.bg}
\affiliation{Bulgarian Academy of Sciences, INRNE} 

\begin{abstract}
 We present a systematic comparison of statistical approaches to Baryon Acoustic Oscillation (BAO) analysis using DESI DR2 data. We evaluate four methods for handling the nuisance parameter $\beta=1/(H_0 r_d)$: marginalization, profiling, Taylor expansion, and full likelihood analysis across multiple cosmological models. Our results demonstrate that while these methods yield consistent constraints for $\Lambda$CDM and $\Omega_K$CDM models, they produce notable differences for models with dynamical dark energy parameters. Through eigenvalue decomposition of Fisher matrices, we identify extreme parameter degeneracies in $ww_a$CDM and $\Omega_Kww_a$CDM models that explain these statistical sensitivities. Surprisingly, $\Omega_K$CDM shows the highest information content across datasets, suggesting BAO measurements are particularly informative about spatial curvature. We further use skewness and kurtosis analysis to identify deviations from Gaussianity, highlighting limitations in Fisher approximations in the dark energy models. Our analysis demonstrates the importance of careful statistical treatment when extracting cosmological constraints from increasingly precise measurements.
\end{abstract}

\maketitle

\section{Introduction}

Baryon Acoustic Oscillations (BAO) have emerged in recent years as one of the most exciting probes of cosmic expansion history. These ``standard rulers'' encoded in the large-scale structure of the universe allow precise measurements of the cosmic distance scale at various redshifts \cite{Seo:2005ys}. Recent surveys like BOSS, eBOSS, SDSS, WiggleZ \cite{SDSS:2005xqv, SDSS:2008tqn, Kazin:2014qga, BOSS:2016wmc, eBOSS:2020yzd, DES:2025bxy} have significantly improved the precision of BAO measurements, yielding increasingly stringent constraints on cosmological models. The most recent and unprecedented in its precision and volume is the Dark Energy Spectroscopic Instrument  (DESI) \cite{DESI:2024hhd} DR2 which when combined with Cosmic Microwave Background (CMB) data measurements by Planck \cite{Planck:2018vyg} and type Ia supernovae data measurement by Pantheon + \cite{Brout:2022vxf} has confirmed the the preference for dynamical dark energy as alternative of the concordance $\Lambda$CDM model.

This results outlines perfectly one of the biggest problem of cosmology -- the tensions observed in the Hubble parameter $H_0$ in the measurements from the early Universe (CMB) and the late universe (SN), also in the structure formation ($\sigma_8$ tension) and the potential tension in the matter density $\Omega_m$  \cite{DiValentino:2019qzk, Zhao:2017cud, Planck:2018vyg, Bahamonde:2021gfp, Abdalla:2022yfr, DiValentino:2025sru}. 

The choice of statistical methodology becomes particularly important when examining tensions in cosmology, since different statistical treatments might amplify or reduce apparent tensions between datasets, potentially leading to different conclusions about the need for physics beyond the standard $\Lambda$CDM model.
In particular, the BAO data do not depend explicitly on the sound horizon at drag epoch ($r_d$) and the Hubble constant ($H_0$), but only on their combination $c/H_0 r_d$. There are different approaches to disentangling this degeneracy. Usually they are taken from other probes such as from CMB or SN, which calibrates the BAO results to the early or late universe \cite{Pogosian:2020ded, Benisty:2020otr,  Staicova:2022zuh, DESI:2024hhd}. Other approaches include a marginalization \cite{Camarena:2019rmj,Nunes:2020hzy, Nunes:2020uex, Staicova:2021ntm}, profile likelihoods \cite{Babic:2022dws, Cuceu:2020dnl, DES:2018fiv, Planck:2013nga} or nonparametric reconstructions \cite{Arjona:2021hmg, Dialektopoulos:2023dhb, Benisty:2022psx}.

In this work, we systematically compare different statistical approaches to BAO likelihood analysis using the latest DESI data. We implement and test four methods: (i) marginalization, which integrates over the parameter space weighted by the prior distribution; (ii) profiling, which maximizes the likelihood at each point in the remaining parameter space; (iii) Taylor expansion approximations of the marginalized likelihood; or (iv) full likelihood.  While these methods are mathematically well-defined, their practical implementations and impacts on cosmological inference have not been systematically compared in the context of recent BAO datasets.  We examine how these methodological choices impact cosmological parameter constraints, particularly on $\Omega_m$ and the dark energy equation of state parameters $w_0$ and $w_a$. In addition to comparing different likelihood estimation methods, we also quantify the non-Gaussianity of the posterior distributions by computing the skewness and kurtosis of marginalized parameters and we use the Fisher information matrix to study degeneracies. This allows us to systematically assess when and why Gaussian approximations break down.

Our paper is organized as follows: In Section~\ref{sec:methods}, we present the cosmological background, while in Section ~\ref{sec:approaches} we present the mathematical formulation of the different statistical approaches. Section~\ref{sec:results} compares the resulting parameter constraints and discusses the implications of methodological choices. We conclude and discuss future prospects in Section~\ref{sec:discussion}.

\section{Measuring cosmological distances}
\label{sec:methods}

\subsection{BAO Likelihood Formulation}
\label{subsec:likelihood}

BAO measurements provide constraints on the comoving angular diameter distance $D_M(z)/r_d$ and the Hubble distance $D_H(z)/r_d = c/(H(z)r_d)$ at various redshifts $z$, where $r_d$ is the sound horizon at radiation drag. Some measurements are reported as the volume-averaged distance $
D_V(z)/r_d = \left[z D_M(z)^2 D_H(z)\right]^{1/3}/r_d $.

The theoretical predictions for these distances depend on the cosmological parameters $\boldsymbol{\theta} = \{\Omega_k, \Omega_m, w_0, w_a, ...\}$ through the expansion history. For a given cosmological model, the dimensionless Hubble parameter $E(z) = H(z)/H_0= \sqrt{\Omega_m(1+z)^3 + \Omega_k(1+z)^2 + \Omega_{DE}(z)}
$. For the dark energy density $\Omega_{DE}(z)$ we use  the CPL parameterization \citep{Chevallier:2000qy, Linder:2005ne} is:

\begin{equation}
\Omega_{DE}(z) = \Omega_{\Lambda}(1+z)^{3(1+w_0+w_a)}e^{-3w_a z/(1+z)}
\end{equation}

where $\Omega_\Lambda= (1-\Omega_m-\Omega_k)$. The comoving distance is then:

\begin{equation}
\chi(z) = \int_0^z \frac{dz'}{E(z')}.
\end{equation}

From these basic quantities, we compute the theoretical predictions from $D_M(z) = \frac{c}{H_0} S_k(\chi(z))$ and 

\begin{equation}
S_k(\chi) = 
\begin{cases}
\sin(\sqrt{-\Omega_k}\chi)/\sqrt{-\Omega_k} & \text{for}~\Omega_k < 0 \\
\chi & \text{for}~\Omega_k = 0 \\
\sinh(\sqrt{\Omega_k}\chi)/\sqrt{\Omega_k} & \text{for}~\Omega_k > 0
\end{cases}
\end{equation}

The BAO likelihood is then constructed by comparing theoretical predictions to measured values as $\mathcal{L}(\boldsymbol{\theta}, \beta) \propto \exp\left(-\frac{1}{2}\chi^2(\boldsymbol{\theta}, \beta)\right)$, 
where and the chi-squared is:

\begin{equation}
\chi^2(\boldsymbol{\theta}, \beta) = \sum_{i,j} [d_i - t_i(\boldsymbol{\theta}, \beta)] C^{-1}_{ij} [d_j - t_j(\boldsymbol{\theta}, \beta)].
\end{equation}

Here, $d_i$ represents the BAO measurements, $t_i(\boldsymbol{\theta}, \beta)$ are the corresponding theoretical predictions, and $C^{-1}_{ij}$ is the inverse covariance matrix. The degenerate parameters $H_0, r_d$ enter only as $\beta = 1/(H_0 r_d)$. 

\subsection{Supernova Likelihood}
\label{subsec:sn_likelihood}

Type Ia supernovae (SNe Ia) serve as standardizable candles, providing distance measurements across a wide range of redshifts. For it, we use the marginalized form, which eliminates $H_0$ and $M_B$ and alows us to use it directly in the marginalized BAO approach. This way we replaced the standard SN likelihood:

with the marginalized one:

\begin{equation}
\chi^2_{\text{SN,marg}}(\boldsymbol{\theta}) = A(\boldsymbol{\theta}) - \frac{B(\boldsymbol{\theta})^2}{C} + \ln\left(\frac{C}{2\pi}\right)
\end{equation}

where
\begin{align}
A(\boldsymbol{\theta}) &= \sum_{i,j} [\mu_{\text{obs},i} - \mu_{\text{th},i}(\boldsymbol{\theta})] C^{-1}_{ij} [\mu_{\text{obs},j} - \mu_{\text{th},j}(\boldsymbol{\theta})] \\
B(\boldsymbol{\theta}) &= \sum_{i,j} [\mu_{\text{obs},i} - \mu_{\text{th},i}(\boldsymbol{\theta})] C^{-1}_{ij} \\
C &= \sum_{i,j} C^{-1}_{ij}.
\end{align}

Here $\mu_{\text{obs}}(z) = m_B(z) - M_B$ is the observed distance modulus, where $m_B$ is the observed apparent magnitude and $M_B$ is the absolute magnitude. The theoretical prediction, $\mu_{\text{th}}(z) = 5\log_{10}\left[\frac{(1+z)c}{H_0}\chi(z)\right] + 25$. $\mathbf{C}$ the covariance matrix that includes both statistical and systematic uncertainties.

This marginalized likelihood depends only on the cosmological parameters $\boldsymbol{\theta}$ through the dimensionless comoving distance $\chi(z)$, and is independent of both $M_B$ and $H_0$. 

\section{Statistical approaches}
\label{sec:approaches}

In this section we describe all the statistical methods employed in this paper - the four likelihoods approaches to handling the nuisance parameter $\beta$ and the post-processing methods we use 
--  the Effective Likelihood and the Asymptotic Likelihood using the Fisher matrix. 
\subsection{Likelihoods}
\subsubsection{Full Likelihood}
\label{subsubsec:full}

The full likelihood approach explicitly includes $H_0$ and $r_d$ as parameters in the model, rather than working with the combination $\beta = 1/(H_0 r_d)$. The chi-squared is calculated as:

\begin{equation}
\chi^2_\text{full}(\boldsymbol{\theta}, H_0, r_d) = \sum_{i,j} [d_i - t_i(\boldsymbol{\theta}, H_0, r_d)] C^{-1}_{ij} [d_j - t_j(\boldsymbol{\theta}, H_0, r_d)]
\end{equation}

This approach requires explicit priors on both $H_0$ and $r_d$. 
The theoretical predictions depend on the parameter $\beta$ through:

\begin{align}
\frac{D_M(z)}{r_d} &= c\beta \cdot g(\boldsymbol{\theta}, z)  \label{f_g}\\
\frac{D_H(z)}{r_d} &= c\beta \cdot f(\boldsymbol{\theta}, z) \nonumber \\
\frac{D_V(z)}{r_d} &= c\beta \cdot [z \cdot g(\boldsymbol{\theta}, z)^2 \cdot f(\boldsymbol{\theta}, z)]^{1/3} \nonumber
\end{align}

where $g(\boldsymbol{\theta}, z) = S_k(\chi(z))$ and $f(\boldsymbol{\theta}, z) = 1/E(z)$ are dimensionless functions.

\subsubsection{Marginalized Likelihood}
\label{subsubsec:marg}

The marginalized likelihood integrates over the nuisance parameter $\beta$ to obtain:
$\mathcal{L}_\text{marg}(\boldsymbol{\theta}) = \int \mathcal{L}(\boldsymbol{\theta}, \beta) \pi(\beta) d\beta$
where $\pi(\beta)$ is the prior on $\beta$. 
For BAO data, the theoretical predictions are linear in $\beta$, allowing the chi-squared to be written as:

\begin{equation}
\chi^2(\boldsymbol{\theta}, \beta) = C - B(\boldsymbol{\theta})\beta + A(\boldsymbol{\theta})\beta^2
\end{equation}

where:
\begin{align}
A(\boldsymbol{\theta}) &= \sum_{i,j} p_i(\boldsymbol{\theta}) C^{-1}_{ij} p_j(\boldsymbol{\theta}) \\
B(\boldsymbol{\theta}) &= \sum_{i,j} d_i C^{-1}_{ij} p_j(\boldsymbol{\theta}) + \sum_{i,j} p_i(\boldsymbol{\theta}) C^{-1}_{ij} d_j \\
C &= \sum_{i,j} d_i C^{-1}_{ij} d_j
\end{align}

with $p_i(\boldsymbol{\theta})$ being the theoretical prediction without the $\beta$ factor (i.e. Eqs. \ref{f_g}). The marginalized chi-squared is then analytically derived as:

\begin{equation}
\chi^2_\text{marg}(\boldsymbol{\theta}) = C - \frac{B(\boldsymbol{\theta})^2}{4A(\boldsymbol{\theta})} + \ln\left(\frac{A(\boldsymbol{\theta})}{2\pi}\right)
\end{equation}

\subsubsection{Profile Likelihood}
\label{subsubsec:profile}
The profile likelihood approach maximizes the likelihood with respect to the nuisance parameter at each point in the remaining parameter space:

\begin{equation}
\mathcal{L}_{\mathrm{prof}}(\boldsymbol{\theta}) = \max_{\beta} \mathcal{L}(\boldsymbol{\theta}, \beta)
\end{equation}

In our implementation, we directly minimize the chi-squared function with respect to $\beta$ for each fixed value of the cosmological parameters $\boldsymbol{\theta}$:

\begin{equation}
\chi^2_{\mathrm{prof}}(\boldsymbol{\theta}) = \min_{\beta} \chi^2(\boldsymbol{\theta}, \beta)
\end{equation}

This minimization is performed numerically using a scalar optimization algorithm. The profile likelihood does not account for the volume of the parameter space associated with the nuisance parameter, which can lead to differences compared to marginalization when parameter degeneracies are present.

\subsubsection{Taylor Expansion Approximation}

The Taylor expansion approach directly approximates the marginalized likelihood by expanding the integrand around the maximum of the likelihood with respect to $\beta$ and then performing the resulting Gaussian integral analytically. Starting with:

\begin{equation}
\mathcal{L}_{\mathrm{marg}}(\boldsymbol{\theta}) = \int \mathcal{L}(\boldsymbol{\theta}, \beta)\pi(\beta)d\beta
\end{equation}

We expand the log-likelihood around $\beta_{\mathrm{opt}}(\boldsymbol{\theta})$, which is the value that maximizes the likelihood:

\begin{equation}
\ln \mathcal{L}(\boldsymbol{\theta}, \beta) \approx \ln \mathcal{L}(\boldsymbol{\theta}, \beta_{\mathrm{opt}}) - \frac{1}{2}A(\boldsymbol{\theta})(\beta - \beta_{\mathrm{opt}})^2
\end{equation}

where $\beta_{\mathrm{opt}} = B(\boldsymbol{\theta})/A(\boldsymbol{\theta})$ and $A(\boldsymbol{\theta})$ represents the curvature of the likelihood at the maximum. With this quadratic approximation and assuming a flat prior on $\beta$, the integral becomes a Gaussian integral that can be solved analytically:

\begin{equation}
\mathcal{L}_{\mathrm{taylor}}(\boldsymbol{\theta}) \approx \mathcal{L}(\boldsymbol{\theta}, \beta_{\mathrm{opt}})\sqrt{\frac{2\pi}{A(\boldsymbol{\theta})}}
\end{equation}

The corresponding chi-squared is:

\begin{equation}
\chi^2_{\mathrm{taylor}}(\boldsymbol{\theta}) = C - \frac{B(\boldsymbol{\theta})^2}{A(\boldsymbol{\theta})} - \ln\left(\frac{2\pi}{A(\boldsymbol{\theta})}\right)
\end{equation}

While this expression seems similar to the Marginalized form, it uses the maximum of the likelihood with respect to $\beta$ thus in a way serving as a connection between the profile and the marginalized form.

\subsection{Approximate Bayesian Computation (ABC)}
We complement our likelihood-based analysis with Approximate Bayesian Computation (ABC), a likelihood-free inference method that directly compares simulated and observed data as measured by a distance metric. 

ABC approximates the standard Bayesian likelihood  posterior by accepting parameter values $\theta$ that produce simulated data $D_{\text{sim}}$ sufficiently similar to observations $D_{\text{obs}}$, as measured by a distance metric $\rho(D_{\text{sim}}, D_{\text{obs}}) \leq \epsilon$. The ABC posterior can be written as:

\begin{equation}
p_{\text{ABC}}(\theta|D_{\text{obs}})\!\propto \!p(\theta) \!\int \!\mathbf{1}(\rho(D_{\text{sim}}, D_{\text{obs}}) \!\leq \!\epsilon) \, p(D_{\text{sim}}|\theta) dD_{\text{sim}}
\end{equation}

We implement an adaptive ABC algorithm that gradually decreases the acceptance threshold $\epsilon$ as sampling progresses, improving posterior estimation. We explore two distance metrics: (i) a covariance-weighted metric $\rho_{\text{cov}}(D_{\text{sim}}, D_{\text{obs}}) = \sqrt{(D_{\text{sim}} - D_{\text{obs}})^T \Sigma^{-1} (D_{\text{sim}} - D_{\text{obs}})}$, which accounts for correlations between BAO measurements and is analogous to a $\chi^2$ statistic; and (ii) a ratio metric $\rho_{\text{ratio}}(D_{\text{sim}}, D_{\text{obs}}) = \sqrt{\sum_i (R_i^{\text{sim}} - R_i^{\text{obs}})^2}$, where $R_i = D_M/D_H$ at redshift $z_i$, which focuses on the consistency of DM/DH ratios across redshifts and is sensitive to the geometric effects of cosmological parameters.

\subsection{Post-processing}
The idea of the post-processing step is to transform the discrete posterior samples we obtained with different likelihoods into a continuous analytical form that captures the essential statistical properties of the likelihood surface.
\subsubsection{Effective Likelihood}
\label{sec:effective_likelihood}
The Effective Likelihood method \cite{Tierney:1986, Lewis:2002ah, DESI:2024cru} approximates the posterior distribution as a multivariate Gaussian, capturing the mean and covariance of the parameters directly from the MCMC chains. This approach is particularly useful for simplifying complex likelihood surfaces while retaining information about parameter correlations. 

For a given model and dataset, we extract the emcee chains and  then compute the mean \( \bar{\theta} \) and covariance matrix \( \Sigma \) of the parameter vector \( \theta \). The effective likelihood is then defined as:
\[
\mathcal{L}_{\text{eff}}(\theta) \propto \exp\left(-\frac{1}{2} (\theta - \bar{\theta})^T \Sigma^{-1} (\theta - \bar{\theta})\right).
\]
 
Furthermore, we can quantify deviations from Gaussianity in the posterior distributions trough two standard statistical moments: skewness (Skew) and kurtosis (Ku), which can be defined for each parameter component separately $\{\theta_i\}$ as:

\begin{equation}
\text{Skew} =
\frac{\frac{1}{n}\sum_{i=1}^{n}(\theta_i - \bar{\theta})^3}{\left(\frac{1}{n}\sum_{i=1}^{n}(\theta_i - \bar{\theta})^2\right)^{3/2}}, \text{Ku} = \frac{\frac{1}{n}\sum_{i=1}^{n}(\theta_i - \bar{\theta})^4}{\left(\frac{1}{n}\sum_{i=1}^{n}(\theta_i - \bar{\theta})^2\right)^2}
\end{equation}

Skewness measures the asymmetry of a probability distribution about its mean (Gaussian distribution has skewness $= 0$). Kurtosis characterizes the heaviness of the tails of a distribution relative to a Gaussian distribution ($=3$). Significant skewness would suggest asymmetric confidence intervals and potential bias in parameter constraints when using Gaussian approximations. Distribution with kurtosis $>3$ has heavier tails and is more prone to outliers than a Gaussian \cite{Cowan:2010js, Sellentin:2015waz}.

\subsubsection{Asymptotic Likelihood (Fisher Matrix Approach)}
\label{sec:fisher}

The Asymptotic Likelihood method utilizes the Fisher information matrix \cite{Tegmark:1996qt, Heavens:2009nx} to approximate the likelihood in the large-data limit, assuming the posterior becomes Gaussian around the maximum likelihood estimate (MLE). Fisher-based approximations have already been used extensively in cosmology \cite{Heavens:2007ka,Takahashi:2009ty, Vallisneri:2011ts, Wolz:2012sr, Seehars:2014ora,Repp:2015jja, Foroozan:2021zzu, Dumerchat:2022uxh,  Ding:2023ibo, Yahia-Cherif:2020knp}. The Fisher matrix \( F_{ij} \) quantifies the curvature of the log-likelihood:
\[
F_{ij} = -\left\langle \frac{\partial^2 \log \mathcal{L}}{\partial \theta_i \partial \theta_j} \right\rangle \Big|_{\theta = \theta_{\text{MLE}}},
\]
where \( \theta_{\text{MLE}} \) is the MLE, approximated here as the mean of the emcee chain. 

Numerically, we compute \( F_{ij} \) using finite differences of the log-posterior values stored in the chains:

\begin{equation}
\mathbf{C} = \frac{1}{N-1}\sum_{i=1}^N (\boldsymbol{\theta}_i - \bar{\boldsymbol{\theta}})(\boldsymbol{\theta}_i - \bar{\boldsymbol{\theta}})^T.
\end{equation}

The Fisher matrix is then approximated as $\mathbf{F} \approx \mathbf{C}^{-1}$. The inverse of the Fisher matrix, \( F^{-1} \), provides the asymptotic covariance matrix, from which we extract the variance of the parameters $\theta_i$, denoted \( \sigma_{\theta_i, \text{asymp}}^2 \) .
 
Unlike standard posterior plots that focus on visualization (i.e. trough packages like \textit{getdist}), the Fisher approach provides a quantitative framework for comparing the dataset constraining power (trough Fisher's determinant) and parameter correlations and degeneracies (trough eigenvalue analysis). The Fisher approach is particularly valuable for understanding the limiting behavior of complex models and highlighting parameter degeneracies that may not be visible in direct MCMC analyses. This reconstruction method assumes a Gaussian posterior and relies on the large-data limit, making it more accurate for highly constrained datasets like BAO+CMB but potentially less reliable for less-constrained cases like BAO alone.

\subsection{Numerical Implementation}
\label{subsec:numerical}

\begin{figure*}[ht!]
    \centering
\includegraphics[width=1\textwidth]{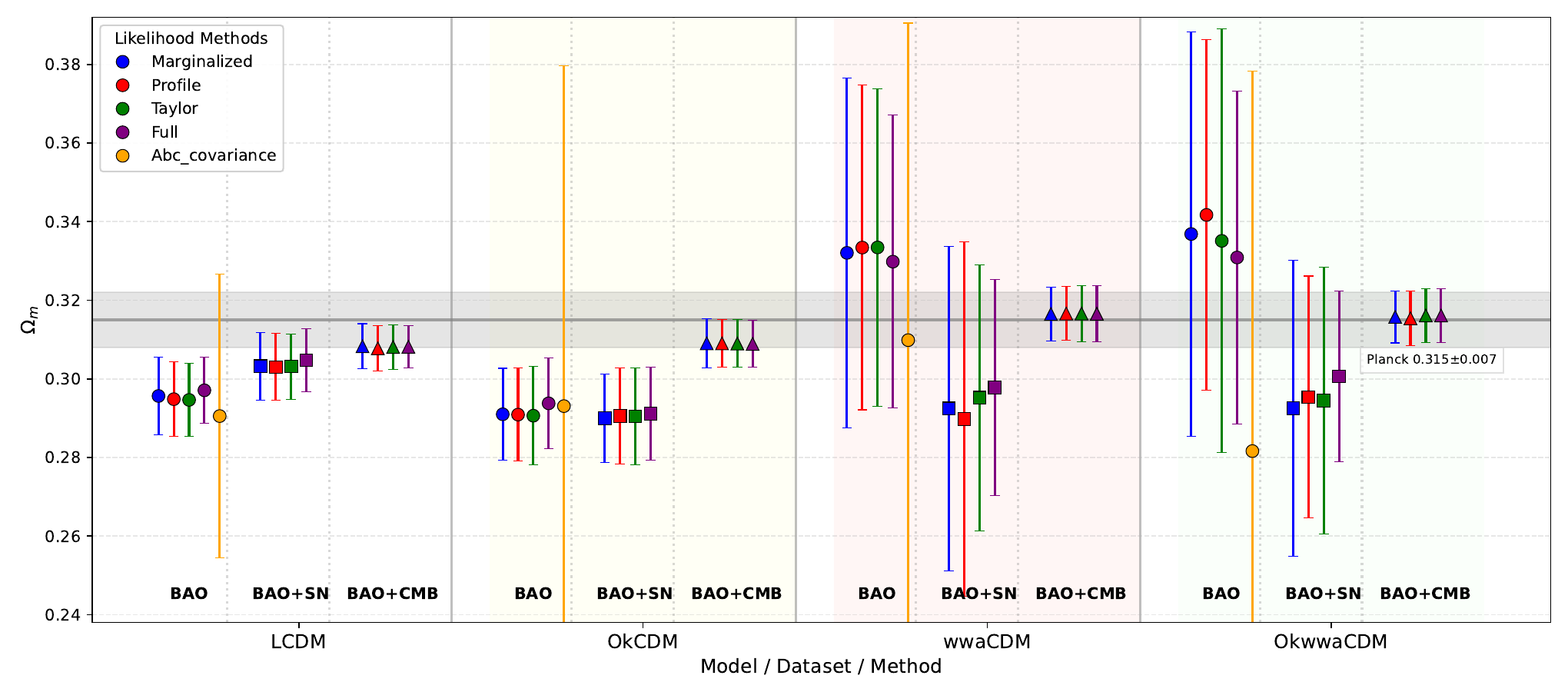} 
 \caption{The summary of the means we obtained for $\Omega_m$ with the different likelihood methods and datasets }
    \label{fig:omega}
\end{figure*}

We implement these four approaches and the related post-processing in a consistent framework to enable direct comparison. For parameter estimation, we employ both Markov Chain Monte Carlo (MCMC) sampling using \texttt{emcee} \footnote{\url{https://github.com/dfm/emcee}}
and nested sampling using \texttt{PolyChord} \citep{Handley:2015fda, Handley:2015vkr}. MCMC provides parameter constraints while nested sampling additionally computes the Bayesian evidence, allowing for model comparison between different statistical approaches.

We use flat uniform priors on for all likelihoods and datasets: $\Omega_k \in [-0.5, 0.5],  \Omega_m \in [0.1, 0.5], w_0 \in [-2.0, 0.0], w_a \in [-2.0, 2.0]$ except for "full" and "BAO+CMB". For the full likelihood approach, we additionally use flat priors on $H_0 \in [60, 80] \text{ km/s/Mpc}, r_d \in [130, 160] \text{ Mpc}$. The CMB prior is fixed by $\Omega_m=0.315 \pm 0.007$ and $\Omega_K=0.001\pm0.004$ when there is curvature.

The priors for ABC are the same as above, but we run two versions: a fixed $H_0 r_d$ one with $H_0=70.0, r_d=147.0$ and version with the full prior on $H_0$ and $r_d$. Due to the large errors associated with the full version, on the plotswe report only the fixed $H_0 r_d$ version. 

The datasets we use are the DESI DR2 \cite{DESI:2024hhd}, and the Pantheon Plus with SH0ES removed \cite{Brout:2022vxf, Benisty:2024lmj}. 

\section{Results}
\label{sec:results}

\subsection{Comparison Across Models and Datasets}

We present the results of our analysis across four cosmological models ($\Lambda$CDM, $\Omega_K$CDM, $ww_a$CDM, and $\Omega_K ww_a$CDM), three data combinations (BAO, BAO+SN, BAO+CMB), and four likelihood approximation methods (marginalized, profile, Taylor expansion, and full). For each combination, we derive constraints on the relevant cosmological parameters $\Omega_m, \Omega_K, w_0, w_a$, focusing on how different statistical approaches and dataset combinations affect the resulting constraints.

\begin{figure}[t!]
    \centering
\includegraphics[width=0.479\textwidth]{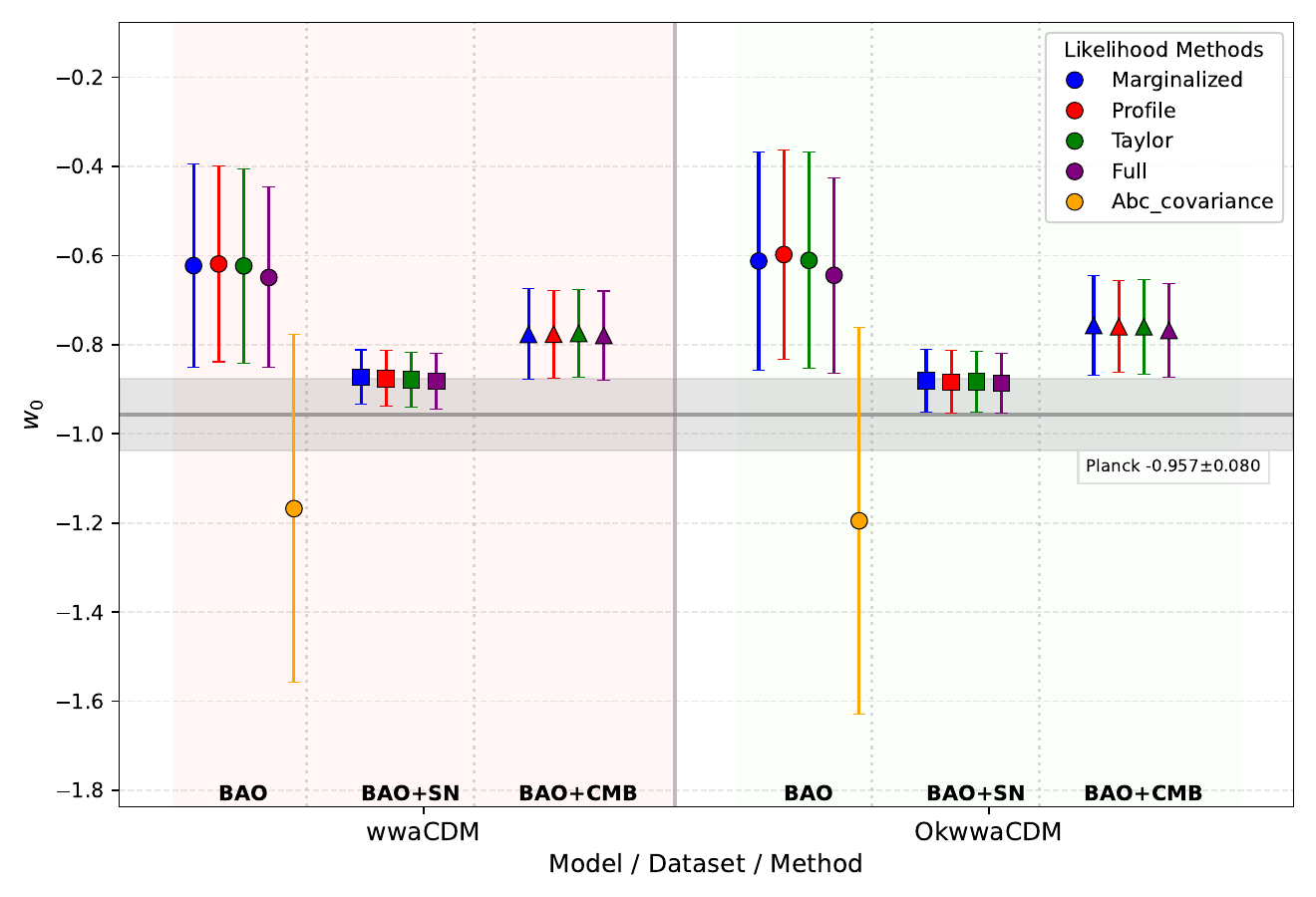} 
\includegraphics[width=0.479\textwidth]{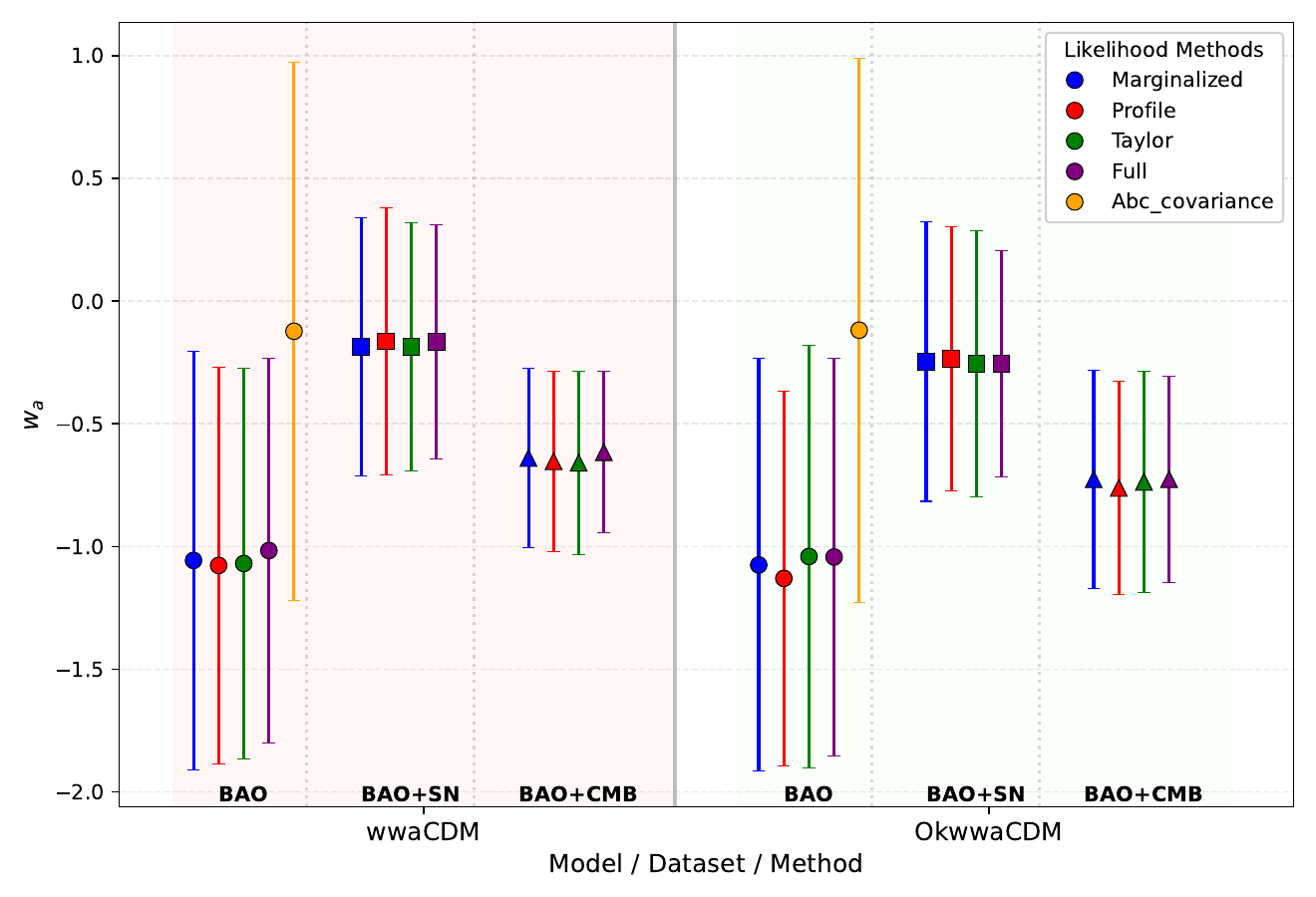} 
\includegraphics[width=0.479\textwidth]{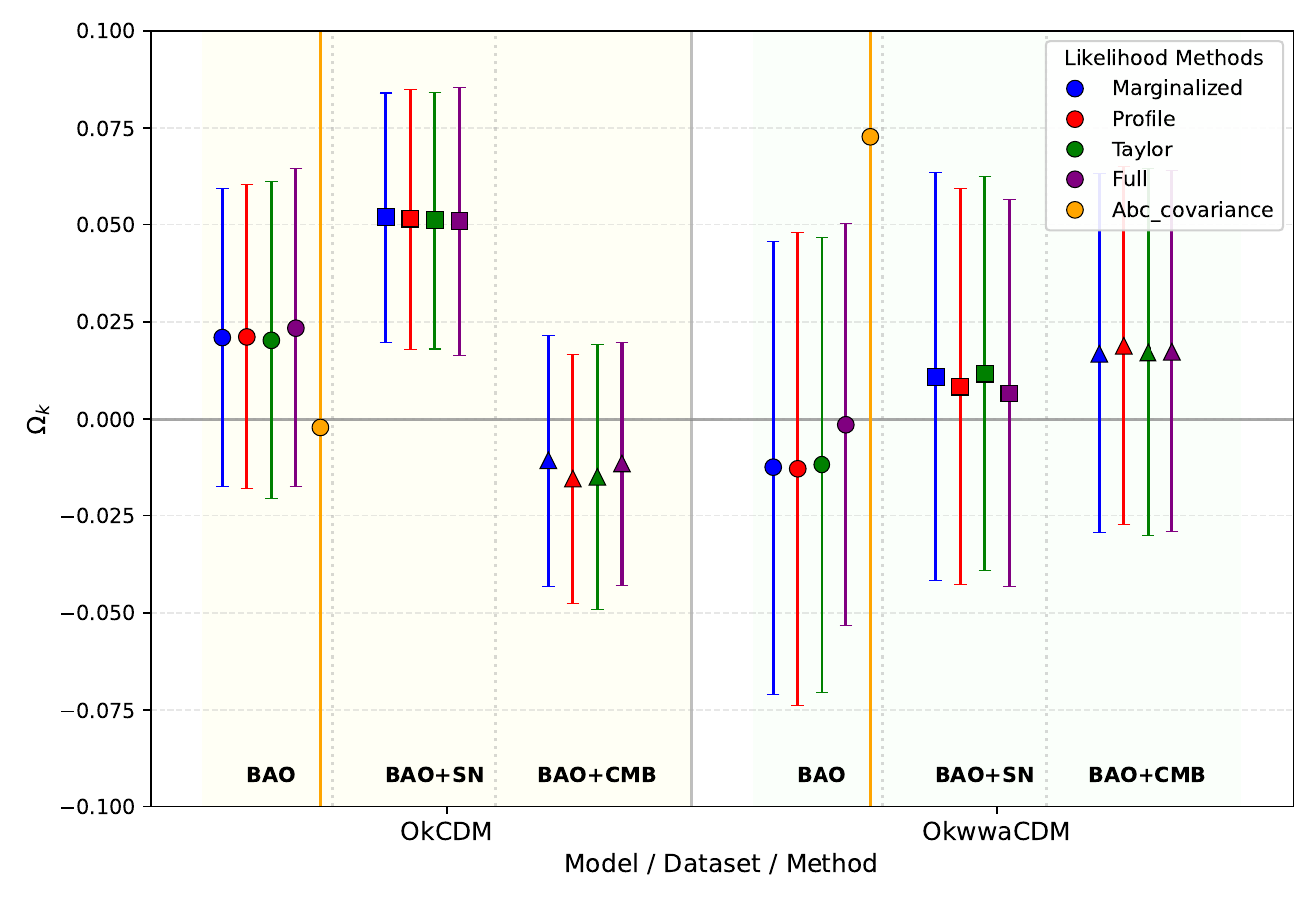} 
 \caption{The summary of the parameters $w_0$ (top), $w_a$ (middle) and $\Omega_K$ (bottom) from the extended models }
    \label{fig:w0wa}
\end{figure}

Figure~\ref{fig:omega} and \ref{fig:w0wa} provides a comprehensive summary of our cosmological constraints across all models, methods, and datasets. The results have been checked with both emcee and polychord, with the full table of the constraints presented in the Appendix \ref{tab:BAO_results}. 

One can see that for $\Lambda$CDM with BAO data, all methods yield $\sim \Omega_m \approx 0.295 \pm 0.01$. In contrast, more complex models ($ww_a$CDM, $\Omega_Kww_a$CDM) show both larger uncertainties and greater sensitivity to the choice of statistical method. The tension between the BAO and the BAO+SN datasets is visible since adding the SN data gives $\Omega_m \approx 0.303 \pm 0.09$. Adding the CMB prior makes all the models to converge to the same $\Omega_m$  ($\approx 0.308 \pm 0.006$), except fro $ww_a$CDM for which we have $\Omega_m\approx 0.316\pm 0.007$. The results are consistent with previously published results, including by the DESI collaboration (for the BAO data for DR2: $\Omega_m = 0.2975 \pm 0.0086$, BAO+CMB $\Omega_m=0.3027 \pm 0.0036$). 

From the plot we can also see that the ABC analysis comes with huge errors, making it unreasonable with such priors. Extending $\Lambda$CDM to include curvature or DDE leads to about 3 fold increase in the errors and significant changes in the mean, especially for $ww_a$CDM. One can see that in these cases, we have much higher matter density $\Omega_m\approx 0.334 \pm 0.06$, a tension that seems to be removed when adding SN data. The same effect is observed also for $\Omega_Kww_a$CDM, with bigger errors. This likely reflects parameter degeneracies between $\Omega_m$ and either curvature or dark energy parameters. When SN data is added, the extended models shift toward lower $\Omega_m$ values, converging around $\Omega_m \approx 0.30$, much closer to $\Lambda$CDM. Adding curvature also changes the mean values for the matter density, but not significantly. The tension is bigger between the BAO alone and the BAO+SN dataset. 
\begin{figure*}[ht!]
    \centering
\includegraphics[width=0.47\textwidth]{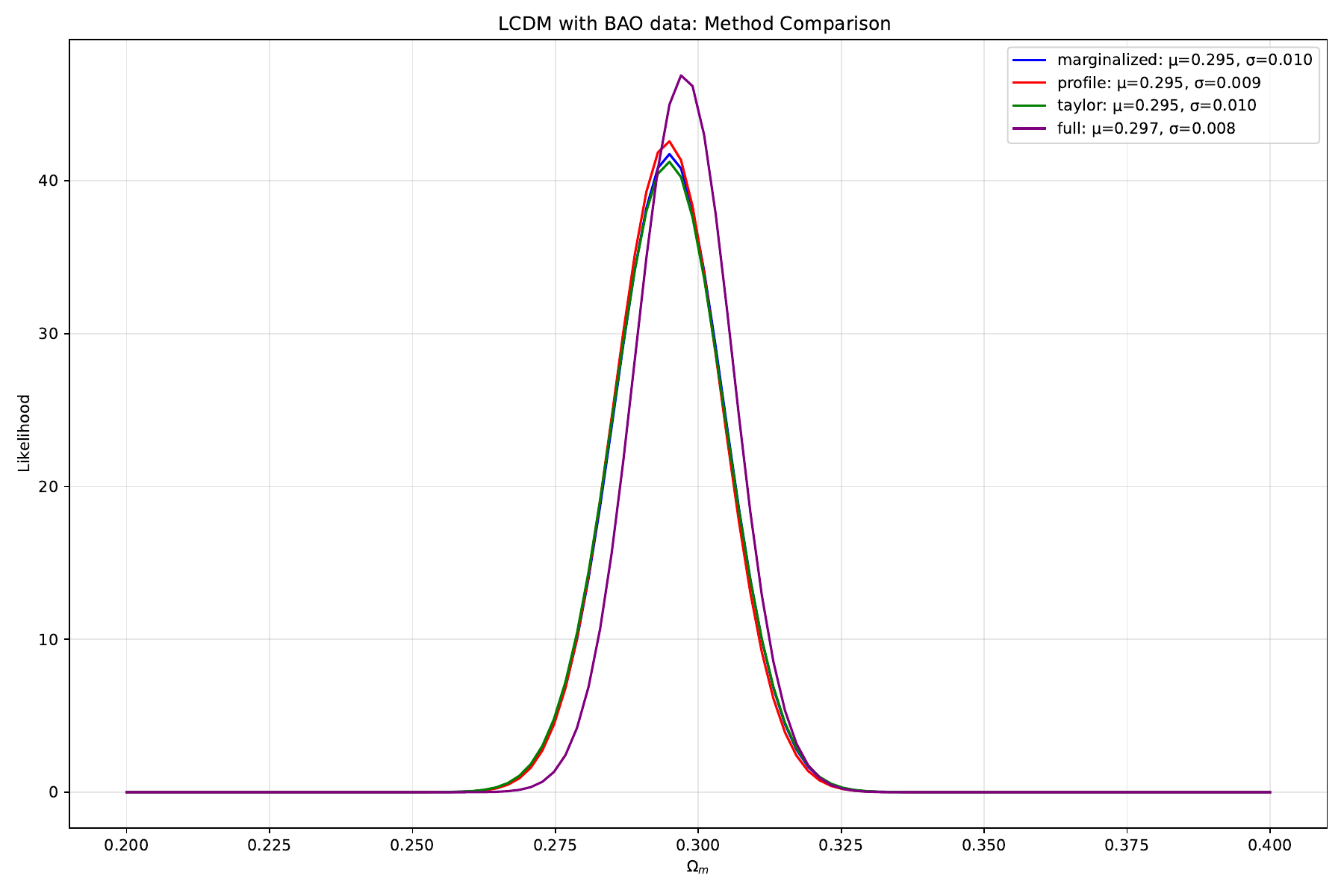} 
\includegraphics[width=0.47\textwidth]{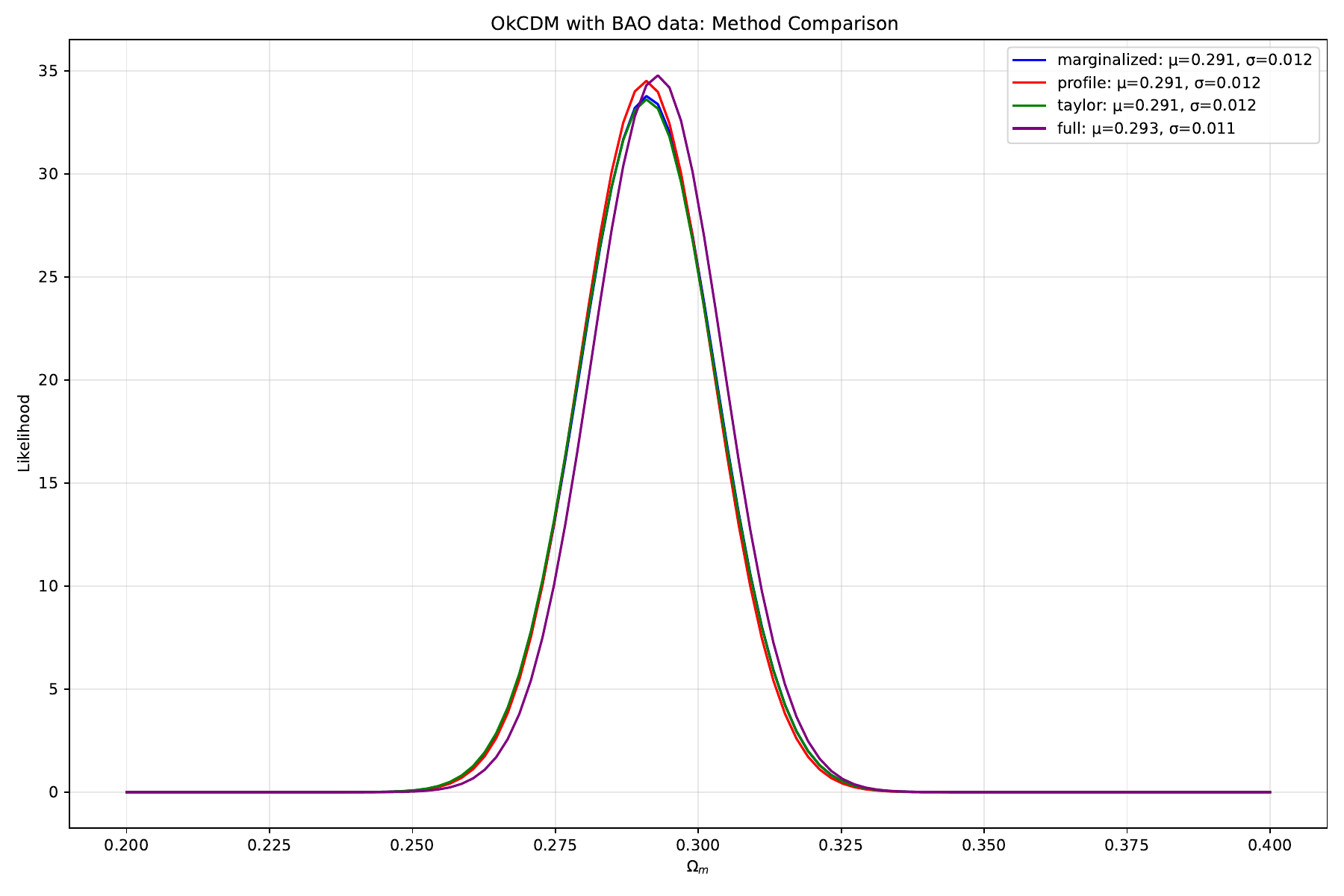} 
\includegraphics[width=0.47\textwidth]{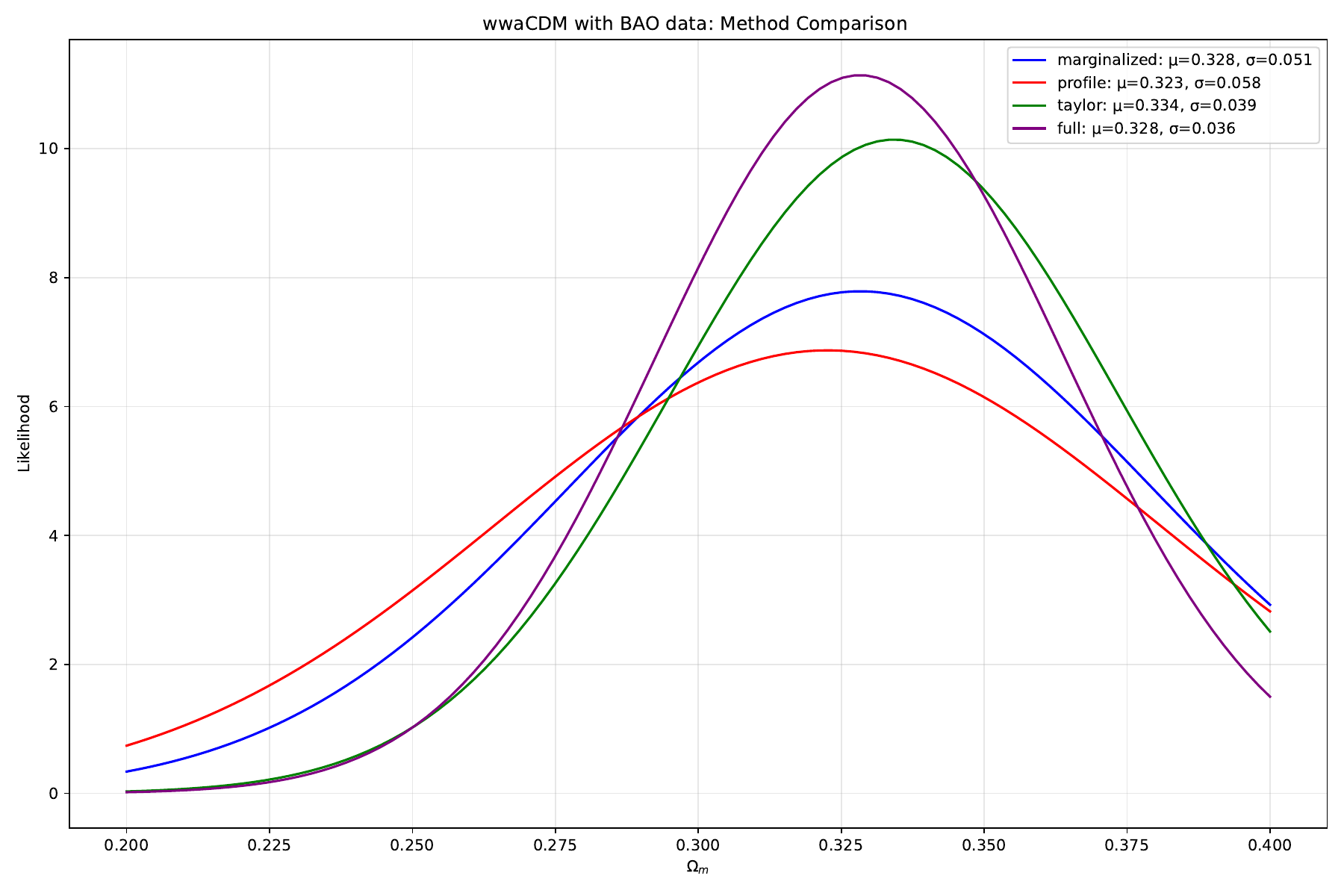} 
\includegraphics[width=0.47\textwidth]{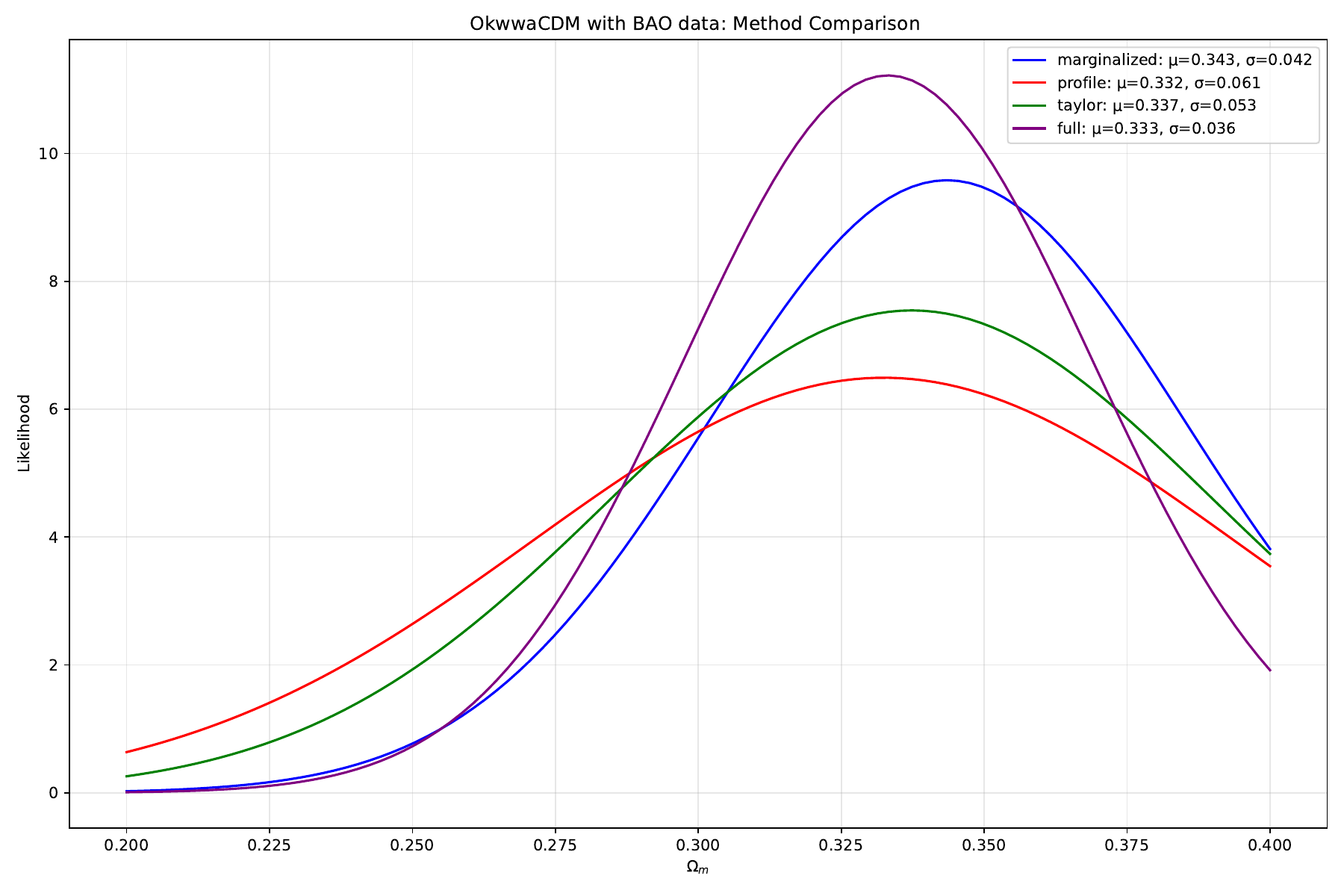} 
 \caption{Effective likelihood method comparison with the BAO dataset for the different models for $\Omega_m$ }
    \label{fig:methods_effective}
\end{figure*}
Analyzing the other parameters, we see that for $\Omega_K$CDM (Fig. \ref{fig:w0wa} bottom), we observe a preference for positive curvature density from BAO alone, and notably, 1 $\sigma$ $\Omega_K>0$ from BAO+SN. For the $\Omega_Kww_a$CDM we see much more neutral results with bigger errors. 

Finally, the possibility for DDE --  (Fig. \ref{fig:w0wa} top), we see that for BAO alone and BAO+CMB, $w_0>-1$ within 1 $\sigma$, which is somewhat mitigated by the addition of SN data. Notably, the very big error prediction of ABC is for $w<-1$. The parameter $w_a$ is weakly constrained, non-zero for the BAO dataset within 1 $\sigma$ but tending to zero for BAO+SN.

\subsection{Method Comparison}

On Fig. \ref{fig:methods_effective}, one can see the effective likelihoods compared for the different models. We see that for the $\Lambda$CDM model, there is a good agreement between the different methods with almost overlapping posteriors Gaussians. The real discrepancies come when adding the DDE model. For the $ww_a$CDM model with BAO data, we find substantial differences between methods (Figure~\ref{fig:BAO_methods_compare}). The Taylor expansion method yields $\Omega_m = 0.334 \pm 0.039$, providing much tighter constraints than marginalized ($\Omega_m = 0.328 \pm 0.051$) or profile ($\Omega_m = 0.323 \pm 0.058$) approaches. 
    
Similar method-dependent effects appear in the $\Omega_Kww_a$CDM model, where the marginalized likelihood gives $\Omega_m = 0.343 \pm 0.042$ while the profile likelihood yields $\Omega_m = 0.332 \pm 0.061$. In general, however, we mostly see good coincidence between the models, with the few differences highlighting the importance of the likelihood choice when analyzing models with parameter degeneracies.

\subsection{Skewedness/kurtosis}

We analyze skewness and kurtosis of our posterior samples to identify when Gaussian approximations may be inadequate. Significant non-Gaussianity in the posteriors (defined as $|\text{Skewness}| > 0.5$ or $|\text{Kurtosis} - 3| > 1$) suggests that full MCMC sampling provides more reliable parameter constraints than analytic approximations that assume Gaussianity.

Our analysis reveals that the $w$CDM and $w_0w_a$CDM models exhibit notable non-Gaussianity in the equation of state parameters, with leptokurtic distributions (kurtosis $>3$) indicating heavier tails than would be predicted by Gaussian approximations. Cosmological likelihoods may exhibit non-Gaussian features due to parameter degeneracies or nonlinear physics, physical boundaries or observational systematics.

\subsection{Fisher vs. MCMC Comparison}

\begin{figure}[t!]
    \centering
\includegraphics[width=0.479\textwidth]{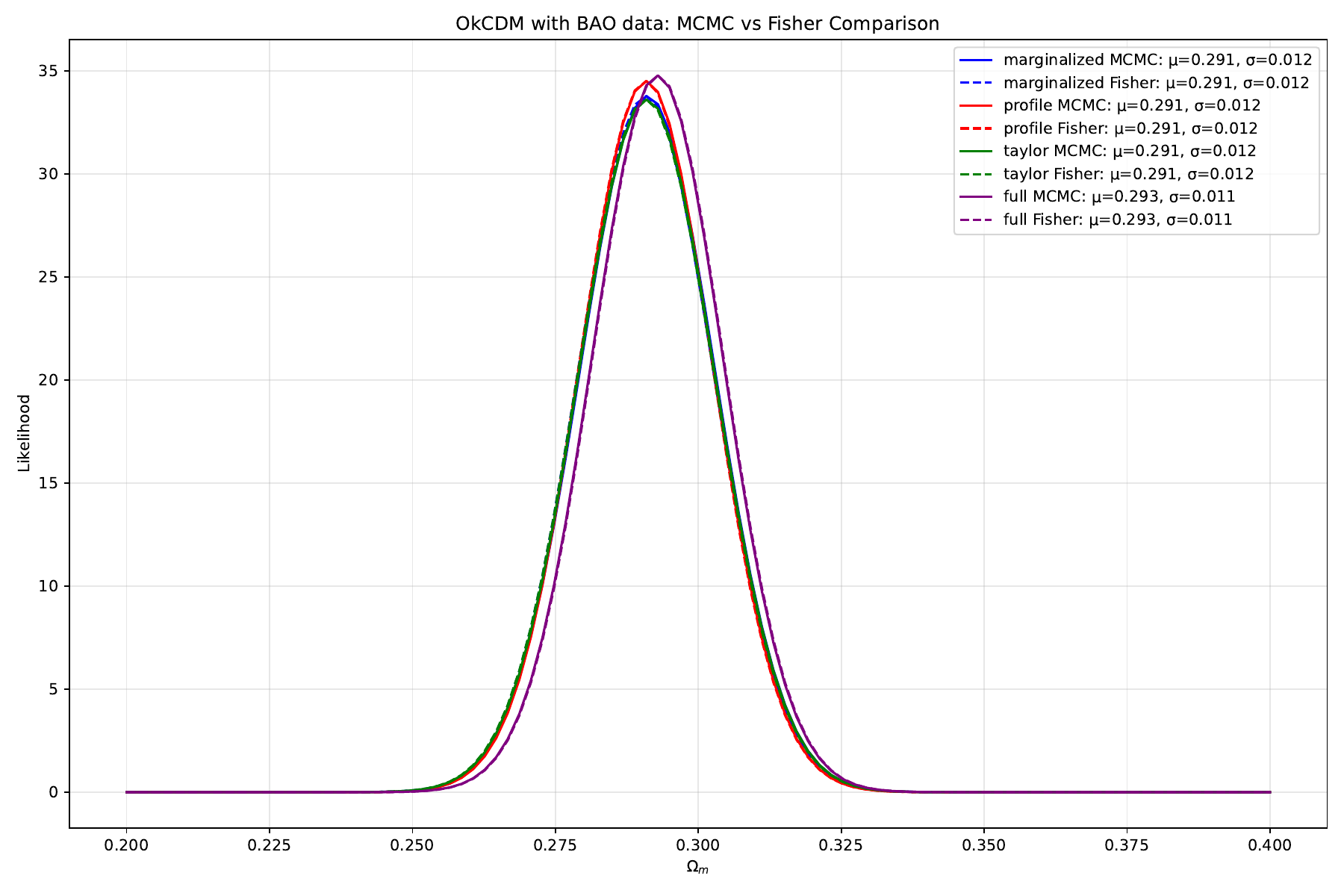} 
\includegraphics[width=0.479\textwidth]{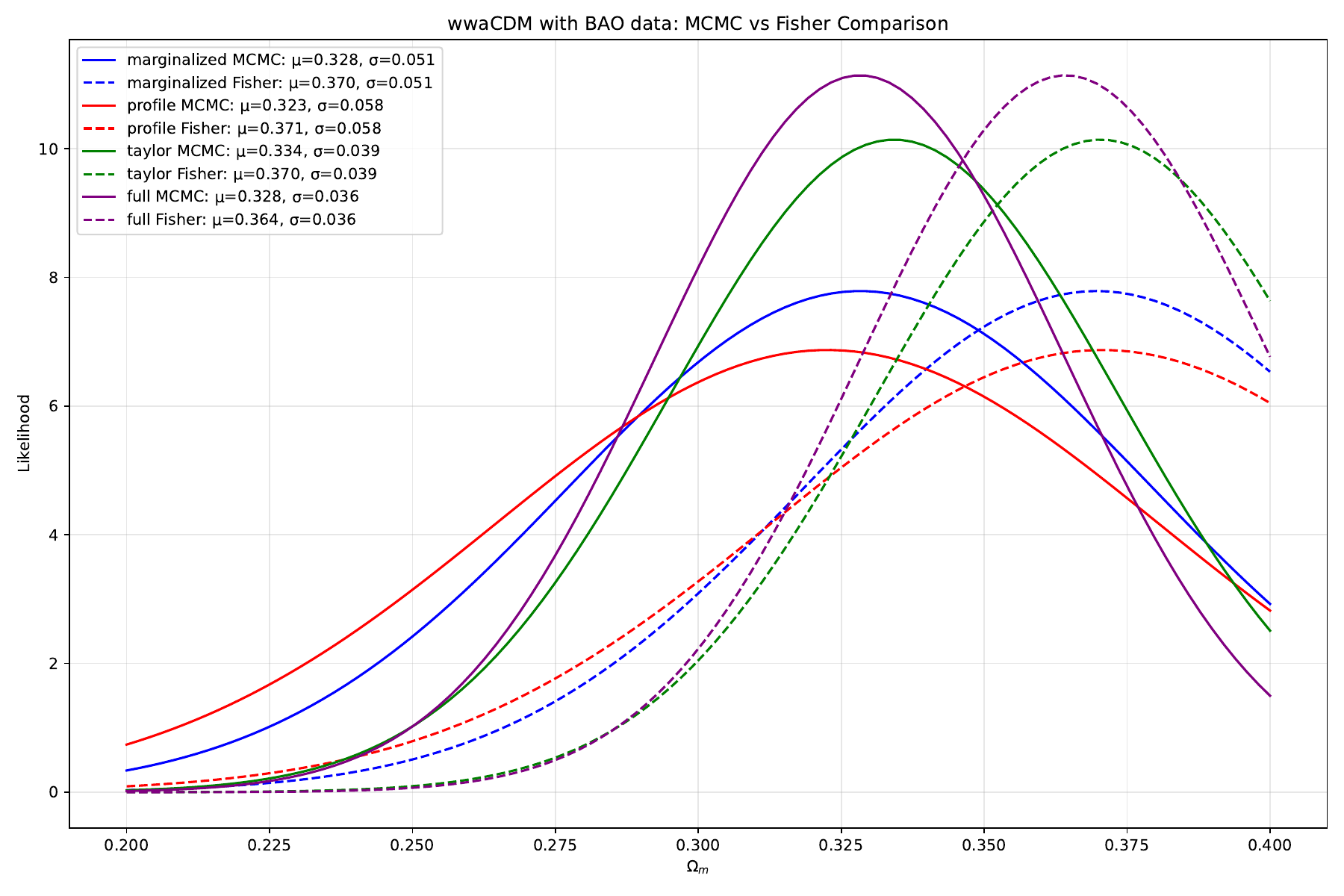} 
 \caption{Comparison between the full MCMC posterior and the Fisher reconstruction for $\Omega_k$CDM -- example of a model where they coincide and $ww_a$CDM - a model where they do not coincide. }
    \label{fig:BAO_methods_compare}
\end{figure}
Our Fisher matrix post-processing of MCMC chains shown on Fig. \ref{fig:BAO_methods_compare} show that for most model-dataset combinations, the Fisher approximation closely matches the full MCMC results, confirming that the posteriors are largely Gaussian. For $ww_a$CDM with BAO data, the Fisher reconstruction deviates from the  the MCMC posteriors, indicating the presence of non-Gaussian features in the likelihood. This is consistent with expectations for models with significant parameter degeneracies when constrained by limited data. The key moment is that the Fisher reconstructions use the maximum likelihood estimate (MLE), which can differ significantly from the posterior mean in non-Gaussian distributions. Therefore, the observed deviations between actual posterior and Fisher reconstruction are indicative for the degeneracies in the model, which might be important in the the light of studying cosmological tensions. 

\subsection{Information content}
\begin{figure}[t!]
    \centering
\includegraphics[width=0.47\textwidth]{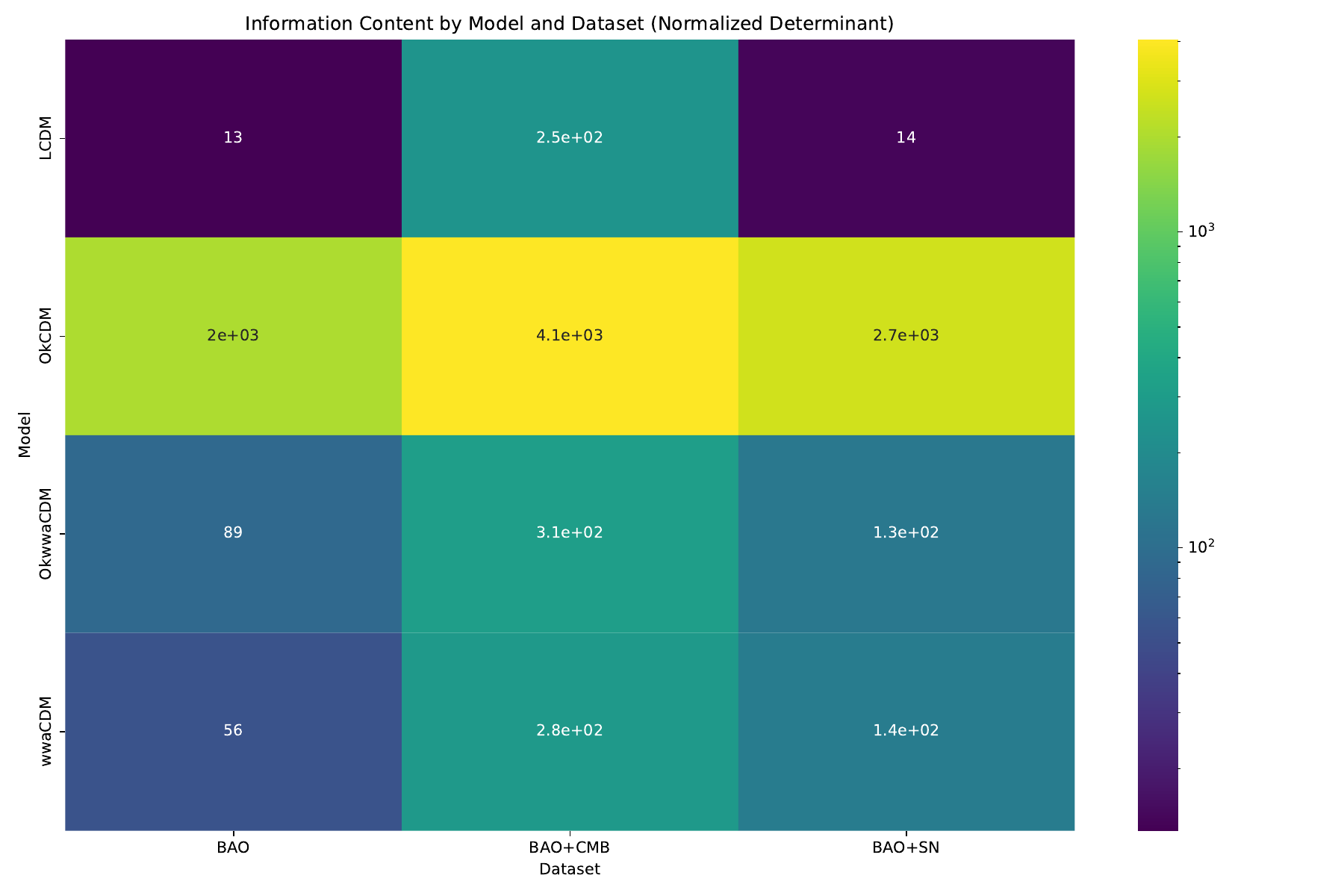} 
\includegraphics[width=0.47\textwidth]{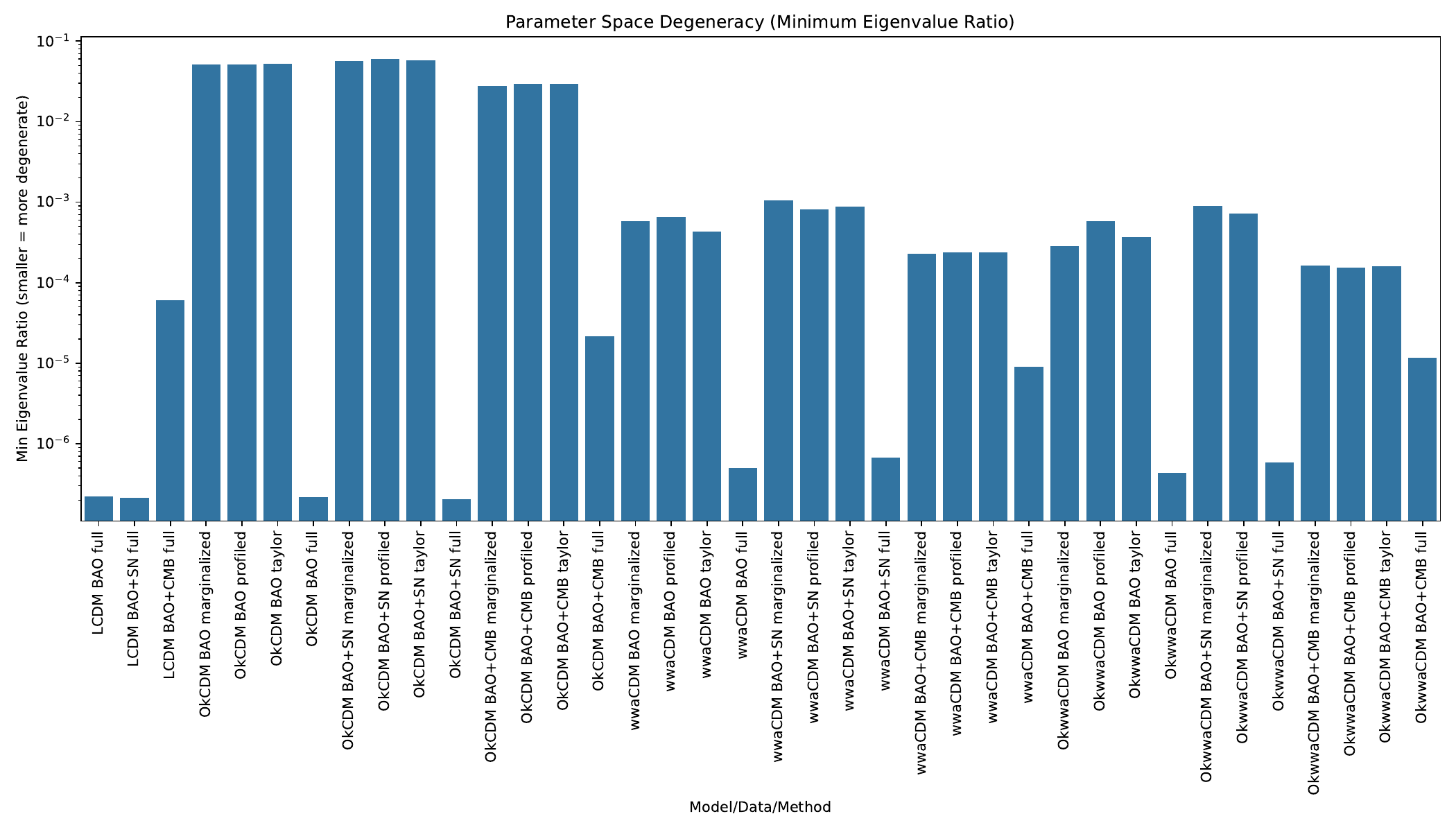} 
 \caption{The information content derived from the Fisher matrix: left the information content of different models depending on the dataset uses and to the right the bar chart with all the models and likelihoods we use}
    \label{fig:information_content}
\end{figure}
We perform an eigenvalue decomposition of the Fisher matrices across all models, datasets, and likelihood methods to analyze parameter space geometries beyond what traditional contour plots reveal. Figure \ref{fig:information_content} presents our findings. The top panel shows the minimum eigenvalue ratio (smallest/largest eigenvalue) on a logarithmic scale, quantifying the degree of parameter degeneracy in each case. $\Lambda$CDM and $\Omega_K$CDM with marginalized, profile, and Taylor methods demonstrate more uniform parameter constraints (ratios $\sim~10^{-2}$), while models with dark energy equation of state parameters exhibit substantially stronger degeneracies (ratios below $10^{-3}$). The bottom panel displays the normalized determinant as a measure of overall information content. 

We see that $\Omega_K$CDM consistently shows the highest information density across all datasets. The result on BAO+CMB combinations having more information comes from the tight CMB prior in these models.  These results highlight how curvature parameters efficiently capture the constraining power of BAO measurements, while dark energy parameters introduce complex degeneracy structures that significantly reduce effective information content despite increasing model flexibility (which can also be seen from the eigenvalues spectrum where specific directions $a w_0+b w_a$ are not well constrained at all). The contrast between parameter degeneracies in different cosmological models demonstrates why Fisher approximations perform adequately for simpler models but break down for $ww_a$CDM and $\Omega_Kww_a$CDM parameter spaces with their highly non-orthogonal constraint directions.

The eigenvalue analysis is relatively robust and independent from the reconstruction methods because it primarily characterizes the local geometry of the likelihood surface around the maximum, not the global shape of the posterior. Therefore the information content metrics and degeneracy measurements should remain valid, even when Fisher approximations of the full posterior fail. When parameter combinations have vastly different constraint strengths (as revealed by the eigenvalue spectrum), the likelihood surface becomes highly anisotropic and non-Gaussian away from the maximum. This makes the quadratic approximation underlying Fisher analysis inadequate for capturing the full posterior shape. Interestingly, we find that the agreement between Fisher and effective likelihoods can \textit{improve} when combining data (e.g., BAO+SN), despite increased skewness in the global distribution hinting that skewness, as a global measure, does not necessarily indicate poor Fisher performance if the \textit{local region} near the peak becomes more symmetric or sharply peaked. 

\subsection{ABC}

Our ABC analysis reveals interesting differences between likelihood-based and likelihood-free inference. For $\Lambda$CDM, the covariance metric yields $\Omega_m = 0.290 \pm 0.032$, consistent with our traditional likelihood analysis, while the ratio metric gives $\Omega_m = 0.312 \pm 0.037$. For models with additional parameters, we find: $\Omega_K$CDM gives $\Omega_m = 0.297 \pm 0.087$ with $\Omega_k = 0.057 \pm 0.109$ (covariance metric); $ww_a$CDM yields $\Omega_m = 0.317 \pm 0.079$, $w_0 = -0.873 \pm 0.385$, and $w_a = 0.154 \pm 1.056$ (covariance metric); and $\Omega_Kww_a$CDM produces similar constraints with slightly larger uncertainties. Notably, the covariance metric produces narrower constraints for simpler models, while models with more parameters show broader distributions, as expected. The ratio metric demonstrates particular sensitivity to curvature in non-flat models, shifting the $\Omega_m$ distribution depending on whether curvature is allowed to vary. This highlights how different distance measures can emphasize different aspects of the cosmological parameter space, potentially revealing degeneracies that might be obscured in traditional likelihood analyses.

\section{Discussion and conclusion} 
\label{sec:discussion}
In this work, we study how our methodological choices affect cosmological parameter constraints by applying different likelihoods to the BAO measurements by DESI DR2. We compare the full BAO likelihood to the marginalized one, the profile one and the Taylor expansion likelihood. We see that while $\Lambda$CDM and $\Omega_K$CDM models show consistency across different likelihood approaches, the more complex dark energy models are more sensitive to the chosen likelihood. 

The eigenvalue decomposition of Fisher matrices shows that these differences emerge due to the dynamical dark energy models ($ww_a$CDM and $\Omega_Kww_a$CDM) exhibiting parameter degeneracies, with directions in the $w_0, w_a$ plane in which the parameters cannot be constrained. These highly anisotropic likelihood surfaces explain why we get varying results depending on the statistical approaches, even for the same datasets. When parameter combinations have different constraint strengths, the likelihood surface becomes highly non-Gaussian away from its maximum, causing the quadratic approximation underlying Fisher analysis to break down. We further examine the non-Gaussianity trough skewness and kurtosis measurements showing that for the DDE models we have leptokurtic distributions with heavy tails that are poorly approximated by Gaussian likelihoods. 

Our finding that $\Omega_K$CDM shows substantially higher information content than either $\Lambda$CDM or the dark energy models is very interesting since it suggests that BAO measurements are more informative about constraining the spatial curvature than when applied to equation of state parameters. 

Our ABC analysis, while yielding larger uncertainties than conventional methods, confirms that when working with weakly constrained models, informative priors become essential for meaningful inference. The large uncertainties in our ABC results without strong priors reflect the inherent limitations of current data when applied to models with higher complexity.

The comparison between datasets shows that adding SN data to BAO can break parameter degeneracies and improve statistical robustness. However, our analysis also shows that even with combined datasets, $ww_a$CDM remains more susceptible to methodological choices than simpler models. The profile likelihood generally handles volume effects less effectively than marginalization because it ignores volume information. The  marginalization integrates over the parameter space, preserving volume effects and it also doesn't rely on Gaussian approximations making it the next best choice after the full likelihood with tight priors.

These findings have important implications for interpreting the tensions in cosmology. Different statistical approaches can yield systematically different results for the same data, particularly for models with significant parameter degeneracies. This underscores the importance of testing multiple statistical methods and examining the likelihood geometry before considering evidence for physics beyond $\Lambda$CDM. It also emphasizes the importance of validating Gaussian approximations before applying Fisher-based forecasts to extended cosmological models, particularly those with poorly constrained or degenerate parameter spaces

In conclusion, our work demonstrates that the statistical foundation of cosmological inference deserves greater attention, particularly as we enter an era of increasingly precise measurements. As data quality improves, the systematic effects of methodological choices may become comparable to or even exceed statistical uncertainties, making a careful statistical approach essential for robust cosmological conclusions.

\section*{Acknowledgement}
This research $w_a$s funded by Bulgarian National Science Fund grant number KP-06-N88/1. 

\section*{Appendix}
Below we present the explicit forms of the effective likelihoods: 

\textbf{Marginalized Effective Likelihood}

Starting with the full MCMC chain in the complete parameter space, we construct a marginalized effective likelihood for $\Omega_m$ by:

\begin{equation}
\mathcal{L}_{\text{eff,marg}}(\Omega_m) \propto \frac{N(\Omega_m)}{\pi(\Omega_m)}
\end{equation}

where $N(\Omega_m)$ is the normalized histogram of $\Omega_m$ values from the MCMC chain, and $\pi(\Omega_m)$ is the prior.

\textbf{Profile Effective Likelihood}

Rather than averaging over nuisance parameters as in marginalization, we profile the posterior samples by:

\begin{equation}
\mathcal{L}_{\text{eff,prof}}(\Omega_m) \propto \max_{\boldsymbol{\phi}} \{p(\Omega_m, \boldsymbol{\phi}|\mathbf{D})/\pi(\Omega_m, \boldsymbol{\phi})\}
\end{equation}

where we identify the maximum posterior value for each narrow bin of $\Omega_m$ and divide by the prior. 

\textbf{Taylor Expansion Effective Likelihood}

To obtain a smooth analytical representation of the likelihood surface, we post-process MCMC samples by fitting a quadratic form to the log-posterior:

\begin{equation}
\ln\mathcal{L}_{\text{eff,taylor}}(\Omega_m) \approx \ln\mathcal{L}_{\text{max}} - \frac{1}{2}\frac{(\Omega_m-\hat{\Omega}_m)^2}{\sigma^2_{\text{eff}}}
\end{equation}

where $\hat{\Omega}_m$ and $\sigma^2_{\text{eff}}$ are determined from the MCMC samples. 

\textbf{Full Effective Likelihood}

For the full effective likelihood, we directly use kernel density estimation on the MCMC samples of $\Omega_m$ (after prior correction):

\begin{equation}
\mathcal{L}_{\text{eff,full}}(\Omega_m) \propto \frac{1}{N}\sum_{i=1}^{N} K\left(\frac{\Omega_m - \Omega_m^{(i)}}{h}\right) / \pi(\Omega_m)
\end{equation}

where $K$ is a kernel function (typically Gaussian) and $h$ is the bandwidth. 

\subsection{Parameters estimates}

\begin{table*}[htbp]
\scriptsize
\centering
\caption{Parameter constraints with BAO data}
\begin{tabular}{llcccc}
\hline
Model & Parameter & marginalized & profile & taylor & full \\
\hline
$\Lambda$CDM & & & & & \\
\hline
 &$ \Omega_m$ (emcee)  & $0.295 \pm 0.010$ & $0.295 \pm 0.009$ & $0.295 \pm 0.010$ & $0.297 \pm 0.008$ \\
 &$ \Omega_m$ (Fisher)  & $0.294 \pm 0.010$ & $0.294 \pm 0.009$ & $0.294 \pm 0.010$ & $0.297 \pm 0.008$ \\
 &$ \Omega_m$ (PolyChord)  & $0.296 \pm 0.045$ & $0.300 \pm 0.051$ & $0.300 \pm 0.047$ & $0.300 \pm 0.047$ \\
 &$ \Omega_m$ (Eff. Skew/Kurt)  & 0.09 / 3.00  & 0.10 / 3.13  & 0.12 / 2.98  & 0.07 / 3.01  \\
\hline\hline
$\Omega_K$CDM & & & & & \\
\hline
 &$ \Omega_m$ (emcee)  & $0.291 \pm 0.012$ & $0.291 \pm 0.012$ & $0.291 \pm 0.012$ & $0.293 \pm 0.011$ \\
 &$ \Omega_m$ (Fisher)  & $0.291 \pm 0.012$ & $0.291 \pm 0.012$ & $0.291 \pm 0.012$ & $0.293 \pm 0.011$ \\
 &$ \Omega_m$ (PolyChord)  & $0.296 \pm 0.044$ & $0.296 \pm 0.052$ & $0.295 \pm 0.046$ & $0.298 \pm 0.047$ \\
 &$ \Omega_m$ (Eff. Skew/Kurt)  & 0.13 / 3.01  & 0.02 / 2.94  & 0.07 / 2.99  & 0.13 / 3.02  \\
 &$ \Omega_k$ (emcee)  & $0.022 \pm 0.040$ & $0.022 \pm 0.040$ & $0.022 \pm 0.040$ & $0.025 \pm 0.040$ \\
 &$ \Omega_k$ (Fisher)  & $0.019 \pm 0.040$ & $0.019 \pm 0.040$ & $0.019 \pm 0.040$ & $0.022 \pm 0.040$ \\
 &$ \Omega_k$ (PolyChord)  & $0.030 \pm 0.122$ & $0.032 \pm 0.136$ & $0.024 \pm 0.122$ & $0.029 \pm 0.136$ \\
 &$ \Omega_k$ (Eff. Skew/Kurt)  & 0.13 / 3.01  & 0.02 / 2.94  & 0.07 / 2.99  & 0.13 / 3.02  \\
\hline\hline
w$w_a$CDM & & & & & \\
\hline
 &$ \Omega_m$ (emcee)  & $0.328 \pm 0.051$ & $0.323 \pm 0.058$ & $0.334 \pm 0.039$ & $0.328 \pm 0.036$ \\
 &$ \Omega_m$ (Fisher)  & $0.370 \pm 0.051$ & $0.371 \pm 0.058$ & $0.370 \pm 0.039$ & $0.364 \pm 0.036$ \\
 &$ \Omega_m$ (PolyChord)  & $0.334 \pm 0.060$ & $0.334 \pm 0.062$ & $0.336 \pm 0.060$ & $0.323 \pm 0.061$ \\
 &$ \Omega_m$ (Eff. Skew/Kurt)  & -2.23 / 8.57  & -2.09 / 7.33  & -2.03 / 9.33  & -1.28 / 4.66  \\
 & $w_0$ (emcee)  & $-0.631 \pm 0.225$ & $-0.647 \pm 0.229$ & $-0.620 \pm 0.219$ & $-0.664 \pm 0.221$ \\
 & $w_0$ (Fisher)  & $-0.355 \pm 0.225$ & $-0.362 \pm 0.229$ & $-0.353 \pm 0.219$ & $-0.393 \pm 0.221$ \\
 & $w_0$ (PolyChord)  & $-0.653 \pm 0.389$ & $-0.642 \pm 0.395$ & $-0.644 \pm 0.393$ & $-0.758 \pm 0.412$ \\
 & $w_0$ (Eff. Skew/Kurt)  & -2.23 / 8.57  & -2.09 / 7.33  & -2.03 / 9.33  & -1.28 / 4.66  \\
 & $w_a$ (emcee)  & $-1.002 \pm 0.868$ & $-0.928 \pm 0.917$ & $-1.079 \pm 0.788$ & $-0.946 \pm 0.832$ \\
 & $w_a$ (Fisher)  & $-1.994 \pm 0.868$ & $-1.989 \pm 0.917$ & $-1.999 \pm 0.788$ & $-1.966 \pm 0.832$ \\
 & $w_a$ (PolyChord)  & $-1.103 \pm 1.042$ & $-1.136 \pm 1.035$ & $-1.155 \pm 1.012$ & $-0.881 \pm 1.089$ \\
 & $w_a$ (Eff. Skew/Kurt)  & -2.23 / 8.57  & -2.09 / 7.33  & -2.03 / 9.33  & -1.28 / 4.66  \\
\hline\hline
$\Omega_Kww_a$CDM & & & & & \\
\hline
 &$ \Omega_m$ (emcee)  & $0.343 \pm 0.042$ & $0.332 \pm 0.061$ & $0.337 \pm 0.053$ & $0.333 \pm 0.036$ \\
 &$ \Omega_m$ (Fisher)  & $0.378 \pm 0.042$ & $0.383 \pm 0.061$ & $0.387 \pm 0.053$ & $0.365 \pm 0.036$ \\
 &$ \Omega_m$ (PolyChord)  & $0.336 \pm 0.069$ & $0.335 \pm 0.069$ & $0.335 \pm 0.071$ & $0.322 \pm 0.066$ \\
 &$ \Omega_m$ (Eff. Skew/Kurt)  & -0.16 / 3.05  & -0.19 / 3.09  & -0.18 / 3.02  & -0.02 / 2.90  \\
 &$ \Omega_k$ (emcee)  & $-0.014 \pm 0.060$ & $-0.014 \pm 0.057$ & $-0.011 \pm 0.061$ & $-0.001 \pm 0.055$ \\
 &$ \Omega_k$ (Fisher)  & $-0.022 \pm 0.060$ & $-0.034 \pm 0.057$ & $-0.036 \pm 0.061$ & $-0.001 \pm 0.055$ \\
 &$ \Omega_k$ (PolyChord)  & $0.004 \pm 0.130$ & $0.005 \pm 0.135$ & $0.002 \pm 0.131$ & $0.025 \pm 0.140$ \\
 &$ \Omega_k$ (Eff. Skew/Kurt)  & -0.16 / 3.05  & -0.19 / 3.09  & -0.18 / 3.02  & -0.02 / 2.90  \\
 & $w_0$ (emcee)  & $-0.591 \pm 0.231$ & $-0.615 \pm 0.246$ & $-0.610 \pm 0.248$ & $-0.643 \pm 0.211$ \\
 & $w_0$ (Fisher)  & $-0.332 \pm 0.231$ & $-0.322 \pm 0.246$ & $-0.307 \pm 0.248$ & $-0.379 \pm 0.211$ \\
 & $w_0$ (PolyChord)  & $-0.646 \pm 0.406$ & $-0.669 \pm 0.419$ & $-0.651 \pm 0.410$ & $-0.742 \pm 0.404$ \\
 & $w_0$ (Eff. Skew/Kurt)  & -0.16 / 3.05  & -0.19 / 3.09  & -0.18 / 3.02  & -0.02 / 2.90  \\
 & $w_a$ (emcee)  & $-1.147 \pm 0.740$ & $-1.015 \pm 0.888$ & $-1.068 \pm 0.848$ & $-1.084 \pm 0.760$ \\
 & $w_a$ (Fisher)  & $-1.984 \pm 0.740$ & $-1.977 \pm 0.888$ & $-1.998 \pm 0.848$ & $-1.960 \pm 0.760$ \\
 & $w_a$ (PolyChord)  & $-1.091 \pm 1.036$ & $-1.024 \pm 1.057$ & $-1.057 \pm 1.053$ & $-0.900 \pm 1.064$ \\
 & $w_a$ (Eff. Skew/Kurt)  & -0.16 / 3.05  & -0.19 / 3.09  & -0.18 / 3.02  & -0.02 / 2.90  \\
\hline\hline
\end{tabular}
\label{tab:BAO_results}
\end{table*}

\begin{table*}[htbp]
\scriptsize
\centering
\caption{Parameter constraints with BAO+SN data}
\begin{tabular}{llcccc}
\hline
Model & Parameter & marginalized & profile & taylor & full \\
\hline
$\Lambda$CDM & & & & & \\
\hline
 &$ \Omega_m$ (emcee)  & $0.304 \pm 0.009$ & $0.303 \pm 0.009$ & $0.303 \pm 0.009$ & $0.304 \pm 0.008$ \\
 &$ \Omega_m$ (Fisher)  & $0.303 \pm 0.009$ & $0.303 \pm 0.009$ & $0.303 \pm 0.009$ & $0.304 \pm 0.008$ \\
 &$ \Omega_m$ (PolyChord)  & $0.306 \pm 0.046$ & $0.303 \pm 0.046$ & $0.305 \pm 0.045$ & $0.308 \pm 0.045$ \\
 &$ \Omega_m$ (Eff. Skew/Kurt)  & 0.08 / 3.02  & 0.08 / 3.06  & 0.12 / 3.03  & 0.13 / 3.11  \\
\hline\hline
$\Omega_K$CDM & & & & & \\
\hline
 &$ \Omega_m$ (emcee)  & $0.290 \pm 0.012$ & $0.290 \pm 0.012$ & $0.291 \pm 0.012$ & $0.291 \pm 0.011$ \\
 &$ \Omega_m$ (Fisher)  & $0.290 \pm 0.012$ & $0.290 \pm 0.012$ & $0.290 \pm 0.012$ & $0.291 \pm 0.011$ \\
 &$ \Omega_m$ (PolyChord)  & $0.294 \pm 0.047$ & $0.297 \pm 0.065$ & $0.295 \pm 0.046$ & $0.294 \pm 0.047$ \\
 &$ \Omega_m$ (Eff. Skew/Kurt)  & -0.00 / 2.94  & 0.05 / 3.09  & 0.05 / 2.97  & 0.02 / 2.91  \\
 &$ \Omega_k$ (emcee)  & $0.053 \pm 0.033$ & $0.051 \pm 0.033$ & $0.051 \pm 0.033$ & $0.051 \pm 0.033$ \\
 &$ \Omega_k$ (Fisher)  & $0.051 \pm 0.033$ & $0.051 \pm 0.033$ & $0.051 \pm 0.033$ & $0.050 \pm 0.033$ \\
 &$ \Omega_k$ (PolyChord)  & $0.049 \pm 0.124$ & $0.048 \pm 0.155$ & $0.054 \pm 0.117$ & $0.053 \pm 0.124$ \\
 &$ \Omega_k$ (Eff. Skew/Kurt)  & -0.00 / 2.94  & 0.05 / 3.09  & 0.05 / 2.97  & 0.02 / 2.91  \\
\hline\hline
w$w_a$CDM & & & & & \\
\hline
 &$ \Omega_m$ (emcee)  & $0.294 \pm 0.040$ & $0.297 \pm 0.030$ & $0.297 \pm 0.034$ & $0.298 \pm 0.027$ \\
 &$ \Omega_m$ (Fisher)  & $0.306 \pm 0.040$ & $0.307 \pm 0.030$ & $0.307 \pm 0.034$ & $0.307 \pm 0.027$ \\
 &$ \Omega_m$ (PolyChord)  & $0.304 \pm 0.054$ & $0.300 \pm 0.055$ & $0.304 \pm 0.053$ & $0.303 \pm 0.050$ \\
 &$ \Omega_m$ (Eff. Skew/Kurt)  & -2.82 / 11.63  & -2.52 / 11.49  & -2.78 / 12.25  & -2.59 / 13.85  \\
 & $w_0$ (emcee)  & $-0.875 \pm 0.062$ & $-0.880 \pm 0.062$ & $-0.877 \pm 0.064$ & $-0.885 \pm 0.061$ \\
 & $w_0$ (Fisher)  & $-0.879 \pm 0.062$ & $-0.880 \pm 0.062$ & $-0.881 \pm 0.064$ & $-0.884 \pm 0.061$ \\
 & $w_0$ (PolyChord)  & $-0.912 \pm 0.227$ & $-0.910 \pm 0.229$ & $-0.909 \pm 0.234$ & $-0.916 \pm 0.236$ \\
 & $w_0$ (Eff. Skew/Kurt)  & -2.82 / 11.63  & -2.52 / 11.49  & -2.78 / 12.25  & -2.59 / 13.85  \\
 & $w_a$ (emcee)  & $-0.197 \pm 0.520$ & $-0.206 \pm 0.504$ & $-0.214 \pm 0.508$ & $-0.170 \pm 0.466$ \\
 & $w_a$ (Fisher)  & $-0.299 \pm 0.520$ & $-0.306 \pm 0.504$ & $-0.303 \pm 0.508$ & $-0.297 \pm 0.466$ \\
 & $w_a$ (PolyChord)  & $-0.231 \pm 0.664$ & $-0.213 \pm 0.681$ & $-0.236 \pm 0.675$ & $-0.229 \pm 0.692$ \\
 & $w_a$ (Eff. Skew/Kurt)  & -2.82 / 11.63  & -2.52 / 11.49  & -2.78 / 12.25  & -2.59 / 13.85  \\
\hline\hline
$\Omega_Kww_a$CDM & & & & & \\
\hline
 &$ \Omega_m$ (emcee)  & $0.289 \pm 0.043$ & $0.286 \pm 0.045$ & $0.012 \pm 0.058$ & $0.298 \pm 0.026$ \\
 &$ \Omega_m$ (Fisher)  & $0.306 \pm 0.043$ & $0.305 \pm 0.045$ & --- & $0.309 \pm 0.026$ \\
 &$ \Omega_m$ (PolyChord)  & $0.295 \pm 0.056$ & $0.295 \pm 0.057$ & $0.296 \pm 0.056$ & $0.295 \pm 0.053$ \\
 &$ \Omega_m$ (Eff. Skew/Kurt)  & -0.07 / 2.95  & 0.01 / 3.05  & 3.25 / 70.17  & 0.08 / 3.06  \\
 &$ \Omega_k$ (emcee)  & $0.008 \pm 0.053$ & $0.008 \pm 0.050$ & $0.001 \pm 0.011$ & $0.008 \pm 0.049$ \\
 &$ \Omega_k$ (Fisher)  & $0.004 \pm 0.053$ & $0.005 \pm 0.050$ & --- & $-0.009 \pm 0.049$ \\
 &$ \Omega_k$ (PolyChord)  & $0.036 \pm 0.139$ & $0.033 \pm 0.141$ & $0.033 \pm 0.139$ & $0.044 \pm 0.139$ \\
 &$ \Omega_k$ (Eff. Skew/Kurt)  & -0.07 / 2.95  & 0.01 / 3.05  & 3.25 / 70.17  & 0.08 / 3.06  \\
 & $w_0$ (emcee)  & $-0.882 \pm 0.071$ & $-0.883 \pm 0.067$ & $-0.034 \pm 0.170$ & $-0.887 \pm 0.070$ \\
 & $w_0$ (Fisher)  & $-0.881 \pm 0.071$ & $-0.887 \pm 0.067$ & --- & $-0.872 \pm 0.070$ \\
 & $w_0$ (PolyChord)  & $-0.928 \pm 0.249$ & $-0.926 \pm 0.255$ & $-0.928 \pm 0.248$ & $-0.954 \pm 0.263$ \\
 & $w_0$ (Eff. Skew/Kurt)  & -0.07 / 2.95  & 0.01 / 3.05  & 3.25 / 70.17  & 0.08 / 3.06  \\
 & $w_a$ (emcee)  & $-0.200 \pm 0.580$ & $-0.159 \pm 0.597$ & $-0.014 \pm 0.120$ & $-0.212 \pm 0.481$ \\
 & $w_a$ (Fisher)  & $-0.293 \pm 0.580$ & $-0.292 \pm 0.597$ & --- & $-0.275 \pm 0.481$ \\
 & $w_a$ (PolyChord)  & $-0.243 \pm 0.708$ & $-0.240 \pm 0.712$ & $-0.254 \pm 0.699$ & $-0.229 \pm 0.720$ \\
 & $w_a$ (Eff. Skew/Kurt)  & -0.07 / 2.95  & 0.01 / 3.05  & 3.25 / 70.17  & 0.08 / 3.06  \\
\hline\hline
\end{tabular}
\label{tab:BAO_SN_results}
\end{table*}

\begin{table*}[htbp]
\scriptsize
\centering
\caption{Parameter constraints with BAO+CMB data}
\begin{tabular}{llcccc}
\hline
Model & Parameter & marginalized & profile & taylor & full \\
\hline
$\Lambda$CDM & & & & & \\
\hline
 &$ \Omega_m$ (emcee)  & $0.308 \pm 0.006$ & $0.308 \pm 0.006$ & $0.308 \pm 0.006$ & $0.308 \pm 0.006$ \\
 &$ \Omega_m$ (Fisher)  & $0.308 \pm 0.006$ & $0.308 \pm 0.006$ & $0.308 \pm 0.006$ & $0.308 \pm 0.006$ \\
 &$ \Omega_m$ (PolyChord)  & $0.309 \pm 0.040$ & $0.310 \pm 0.041$ & $0.310 \pm 0.040$ & $0.310 \pm 0.036$ \\
 &$ \Omega_m$ (Eff. Skew/Kurt)  & 0.03 / 3.03  & 0.00 / 3.04  & 0.04 / 3.07  & 0.02 / 3.05  \\
\hline\hline
$\Omega_K$CDM & & & & & \\
\hline
 &$ \Omega_m$ (emcee)  & $0.309 \pm 0.006$ & $0.309 \pm 0.006$ & $0.309 \pm 0.006$ & $0.309 \pm 0.006$ \\
 &$ \Omega_m$ (Fisher)  & $0.309 \pm 0.006$ & $0.309 \pm 0.006$ & $0.309 \pm 0.006$ & $0.309 \pm 0.006$ \\
 &$ \Omega_m$ (PolyChord)  & $0.310 \pm 0.038$ & $0.311 \pm 0.038$ & $0.310 \pm 0.037$ & $0.311 \pm 0.036$ \\
 &$ \Omega_m$ (Eff. Skew/Kurt)  & 0.10 / 3.04  & 0.11 / 3.05  & 0.12 / 3.03  & 0.10 / 3.06  \\
 &$ \Omega_k$ (emcee)  & $-0.013 \pm 0.034$ & $-0.013 \pm 0.033$ & $-0.013 \pm 0.033$ & $-0.012 \pm 0.032$ \\
 &$ \Omega_k$ (Fisher)  & $-0.015 \pm 0.034$ & $-0.015 \pm 0.033$ & $-0.015 \pm 0.033$ & $-0.015 \pm 0.032$ \\
 &$ \Omega_k$ (PolyChord)  & $0.001 \pm 0.129$ & $-0.006 \pm 0.128$ & $-0.003 \pm 0.122$ & $0.003 \pm 0.141$ \\
 &$ \Omega_k$ (Eff. Skew/Kurt)  & 0.10 / 3.04  & 0.11 / 3.05  & 0.12 / 3.03  & 0.10 / 3.06  \\
\hline\hline
w$w_a$CDM & & & & & \\
\hline
 &$ \Omega_m$ (emcee)  & $0.316 \pm 0.007$ & $0.316 \pm 0.007$ & $0.317 \pm 0.007$ & $0.317 \pm 0.007$ \\
 &$ \Omega_m$ (Fisher)  & $0.316 \pm 0.007$ & $0.316 \pm 0.007$ & $0.316 \pm 0.007$ & $0.317 \pm 0.007$ \\
 &$ \Omega_m$ (PolyChord)  & $0.315 \pm 0.037$ & $0.315 \pm 0.037$ & $0.314 \pm 0.038$ & $0.314 \pm 0.034$ \\
 &$ \Omega_m$ (Eff. Skew/Kurt)  & -0.01 / 2.92  & -0.01 / 2.94  & -0.05 / 2.98  & 0.02 / 3.00  \\
 & $w_0$ (emcee)  & $-0.775 \pm 0.099$ & $-0.781 \pm 0.099$ & $-0.775 \pm 0.100$ & $-0.781 \pm 0.100$ \\
 & $w_0$ (Fisher)  & $-0.781 \pm 0.099$ & $-0.784 \pm 0.099$ & $-0.783 \pm 0.100$ & $-0.775 \pm 0.100$ \\
 & $w_0$ (PolyChord)  & $-0.850 \pm 0.297$ & $-0.857 \pm 0.309$ & $-0.861 \pm 0.295$ & $-0.887 \pm 0.317$ \\
 & $w_0$ (Eff. Skew/Kurt)  & -0.01 / 2.92  & -0.01 / 2.94  & -0.05 / 2.98  & 0.02 / 3.00  \\
 & $w_a$ (emcee)  & $-0.658 \pm 0.365$ & $-0.637 \pm 0.363$ & $-0.641 \pm 0.363$ & $-0.613 \pm 0.327$ \\
 & $w_a$ (Fisher)  & $-0.586 \pm 0.365$ & $-0.572 \pm 0.363$ & $-0.577 \pm 0.363$ & $-0.601 \pm 0.327$ \\
 & $w_a$ (PolyChord)  & $-0.523 \pm 0.629$ & $-0.511 \pm 0.659$ & $-0.527 \pm 0.638$ & $-0.496 \pm 0.700$ \\
 & $w_a$ (Eff. Skew/Kurt)  & -0.01 / 2.92  & -0.01 / 2.94  & -0.05 / 2.98  & 0.02 / 3.00  \\
\hline\hline
$\Omega_Kww_a$CDM & & & & & \\
\hline
 &$ \Omega_m$ (emcee)  & $0.316 \pm 0.007$ & $0.316 \pm 0.007$ & $0.316 \pm 0.007$ & $0.316 \pm 0.007$ \\
 &$ \Omega_m$ (Fisher)  & $0.317 \pm 0.007$ & $0.317 \pm 0.007$ & $0.317 \pm 0.007$ & $0.317 \pm 0.007$ \\
 &$ \Omega_m$ (PolyChord)  & $0.313 \pm 0.038$ & $0.314 \pm 0.038$ & $0.313 \pm 0.038$ & $0.312 \pm 0.035$ \\
 &$ \Omega_m$ (Eff. Skew/Kurt)  & -0.16 / 2.99  & -0.19 / 3.11  & -0.13 / 3.10  & -0.15 / 3.02  \\
 &$ \Omega_k$ (emcee)  & $0.017 \pm 0.046$ & $0.017 \pm 0.046$ & $0.016 \pm 0.046$ & $0.017 \pm 0.047$ \\
 &$ \Omega_k$ (Fisher)  & $0.016 \pm 0.046$ & $0.017 \pm 0.046$ & $0.010 \pm 0.046$ & $0.011 \pm 0.047$ \\
 &$ \Omega_k$ (PolyChord)  & $0.032 \pm 0.130$ & $0.028 \pm 0.129$ & $0.028 \pm 0.128$ & $0.030 \pm 0.134$ \\
 &$ \Omega_k$ (Eff. Skew/Kurt)  & -0.16 / 2.99  & -0.19 / 3.11  & -0.13 / 3.10  & -0.15 / 3.02  \\
 & $w_0$ (emcee)  & $-0.754 \pm 0.102$ & $-0.756 \pm 0.107$ & $-0.759 \pm 0.106$ & $-0.756 \pm 0.104$ \\
 & $w_0$ (Fisher)  & $-0.758 \pm 0.102$ & $-0.757 \pm 0.107$ & $-0.774 \pm 0.106$ & $-0.752 \pm 0.104$ \\
 & $w_0$ (PolyChord)  & $-0.834 \pm 0.307$ & $-0.829 \pm 0.309$ & $-0.829 \pm 0.292$ & $-0.864 \pm 0.326$ \\
 & $w_0$ (Eff. Skew/Kurt)  & -0.16 / 2.99  & -0.19 / 3.11  & -0.13 / 3.10  & -0.15 / 3.02  \\
 & $w_a$ (emcee)  & $-0.755 \pm 0.446$ & $-0.757 \pm 0.461$ & $-0.743 \pm 0.449$ & $-0.739 \pm 0.432$ \\
 & $w_a$ (Fisher)  & $-0.730 \pm 0.446$ & $-0.747 \pm 0.461$ & $-0.682 \pm 0.449$ & $-0.693 \pm 0.432$ \\
 & $w_a$ (PolyChord)  & $-0.596 \pm 0.661$ & $-0.602 \pm 0.692$ & $-0.606 \pm 0.669$ & $-0.559 \pm 0.701$ \\
 & $w_a$ (Eff. Skew/Kurt)  & -0.16 / 2.99  & -0.19 / 3.11  & -0.13 / 3.10  & -0.15 / 3.02  \\
\hline\hline
\end{tabular}
\label{tab:BAO_CMB_results}
\end{table*}

\normalsize

\bibliographystyle{unsrt}
\bibliography{references}

\begin{thebibliography}{10}

\bibitem{Seo:2005ys}
Hee-Jong Seo and Daniel~J. Eisenstein.
\newblock {Baryonic acoustic oscillations in simulated galaxy redshift surveys}.
\newblock {\em Astrophys. J.}, 633:575--588, 2005.

\bibitem{SDSS:2005xqv}
Daniel~J. Eisenstein et~al.
\newblock {Detection of the Baryon Acoustic Peak in the Large-Scale Correlation Function of SDSS Luminous Red Galaxies}.
\newblock {\em Astrophys. J.}, 633:560--574, 2005.

\bibitem{SDSS:2008tqn}
Kevork~N. Abazajian et~al.
\newblock {The Seventh Data Release of the Sloan Digital Sky Survey}.
\newblock {\em Astrophys. J. Suppl.}, 182:543--558, 2009.

\bibitem{Kazin:2014qga}
Eyal~A. Kazin et~al.
\newblock {The WiggleZ Dark Energy Survey: improved distance measurements to z = 1 with reconstruction of the baryonic acoustic feature}.
\newblock {\em Mon. Not. Roy. Astron. Soc.}, 441(4):3524--3542, 2014.

\bibitem{BOSS:2016wmc}
Shadab Alam et~al.
\newblock {The clustering of galaxies in the completed SDSS-III Baryon Oscillation Spectroscopic Survey: cosmological analysis of the DR12 galaxy sample}.
\newblock {\em Mon. Not. Roy. Astron. Soc.}, 470(3):2617--2652, 2017.

\bibitem{eBOSS:2020yzd}
Shadab Alam et~al.
\newblock {Completed SDSS-IV extended Baryon Oscillation Spectroscopic Survey: Cosmological implications from two decades of spectroscopic surveys at the Apache Point Observatory}.
\newblock {\em Phys. Rev. D}, 103(8):083533, 2021.

\bibitem{DES:2025bxy}
T.~M.~C. Abbott et~al.
\newblock {Dark Energy Survey: implications for cosmological expansion models from the final DES Baryon Acoustic Oscillation and Supernova data}.
\newblock 3 2025.

\bibitem{DESI:2024hhd}
A.~G. Adame et~al.
\newblock {DESI 2024 VII: Cosmological Constraints from the Full-Shape Modeling of Clustering Measurements}.
\newblock 11 2024.

\bibitem{Planck:2018vyg}
N.~Aghanim et~al.
\newblock {Planck 2018 results. VI. Cosmological parameters}.
\newblock {\em Astron. Astrophys.}, 641:A6, 2020.
\newblock [Erratum: Astron.Astrophys. 652, C4 (2021)].

\bibitem{Brout:2022vxf}
Dillon Brout et~al.
\newblock {The Pantheon+ Analysis: Cosmological Constraints}.
\newblock {\em Astrophys. J.}, 938(2):110, 2022.

\bibitem{DiValentino:2019qzk}
Eleonora Di~Valentino, Alessandro Melchiorri, and Joseph Silk.
\newblock {Planck evidence for a closed Universe and a possible crisis for cosmology}.
\newblock {\em Nature Astron.}, 4(2):196--203, 2019.

\bibitem{Zhao:2017cud}
Gong-Bo Zhao et~al.
\newblock {Dynamical dark energy in light of the latest observations}.
\newblock {\em Nature Astron.}, 1(9):627--632, 2017.

\bibitem{Bahamonde:2021gfp}
Sebastian Bahamonde, Konstantinos~F. Dialektopoulos, Celia Escamilla-Rivera, Gabriel Farrugia, Viktor Gakis, Martin Hendry, Manuel Hohmann, Jackson Levi~Said, Jurgen Mifsud, and Eleonora Di~Valentino.
\newblock {Teleparallel gravity: from theory to cosmology}.
\newblock {\em Rept. Prog. Phys.}, 86(2):026901, 2023.

\bibitem{Abdalla:2022yfr}
Elcio Abdalla et~al.
\newblock {Cosmology intertwined: A review of the particle physics, astrophysics, and cosmology associated with the cosmological tensions and anomalies}.
\newblock {\em JHEAp}, 34:49--211, 2022.

\bibitem{DiValentino:2025sru}
Eleonora Di~Valentino et~al.
\newblock {The CosmoVerse White Paper: Addressing observational tensions in cosmology with systematics and fundamental physics}.
\newblock 4 2025.

\bibitem{Pogosian:2020ded}
Levon Pogosian, Gong-Bo Zhao, and Karsten Jedamzik.
\newblock {Recombination-independent determination of the sound horizon and the Hubble constant from BAO}.
\newblock {\em Astrophys. J. Lett.}, 904(2):L17, 2020.

\bibitem{Benisty:2020otr}
David Benisty and Denitsa Staicova.
\newblock {Testing late-time cosmic acceleration with uncorrelated baryon acoustic oscillation dataset}.
\newblock {\em Astron. Astrophys.}, 647:A38, 2021.

\bibitem{Staicova:2022zuh}
Denitsa Staicova.
\newblock {DE Models with Combined H$_{0}$ \textperiodcentered{} r$_{d}$ from BAO and CMB Dataset and Friends}.
\newblock {\em Universe}, 8(12):631, 2022.

\bibitem{Camarena:2019rmj}
David Camarena and Valerio Marra.
\newblock {A new method to build the (inverse) distance ladder}.
\newblock {\em Mon. Not. Roy. Astron. Soc.}, 495(3):2630--2644, 2020.

\bibitem{Nunes:2020hzy}
Rafael~C. Nunes, Santosh~K. Yadav, J.~F. Jesus, and Armando Bernui.
\newblock {Cosmological parameter analyses using transversal BAO data}.
\newblock {\em Mon. Not. Roy. Astron. Soc.}, 497(2):2133--2141, 2020.

\bibitem{Nunes:2020uex}
Rafael~C. Nunes and Armando Bernui.
\newblock {BAO signatures in the 2-point angular correlations and the Hubble tension}.
\newblock {\em Eur. Phys. J. C}, 80(11):1025, 2020.

\bibitem{Staicova:2021ntm}
Denitsa Staicova and David Benisty.
\newblock {Constraining the dark energy models using baryon acoustic oscillations: An approach independent of H0 \ensuremath{\cdot} rd}.
\newblock {\em Astron. Astrophys.}, 668:A135, 2022.

\bibitem{Babic:2022dws}
Ivana Babi\'c, Fabian Schmidt, and Beatriz Tucci.
\newblock {BAO scale inference from biased tracers using the EFT likelihood}.
\newblock {\em JCAP}, 08(08):007, 2022.

\bibitem{Cuceu:2020dnl}
Andrei Cuceu, Andreu Font-Ribera, and Benjamin Joachimi.
\newblock {Bayesian methods for fitting Baryon Acoustic Oscillations in the Lyman-$\alpha$ forest}.
\newblock {\em JCAP}, 07:035, 2020.

\bibitem{DES:2018fiv}
K.~C. Chan et~al.
\newblock {BAO from Angular Clustering: Optimization and Mitigation of Theoretical Systematics}.
\newblock {\em Mon. Not. Roy. Astron. Soc.}, 480(3):3031--3051, 2018.

\bibitem{Planck:2013nga}
P.~A.~R. Ade et~al.
\newblock {Planck intermediate results. XVI. Profile likelihoods for cosmological parameters}.
\newblock {\em Astron. Astrophys.}, 566:A54, 2014.

\bibitem{Arjona:2021hmg}
Rub\'en Arjona and Savvas Nesseris.
\newblock {Novel null tests for the spatial curvature and homogeneity of the Universe and their machine learning reconstructions}.
\newblock {\em Phys. Rev. D}, 103(10):103539, 2021.

\bibitem{Dialektopoulos:2023dhb}
Konstantinos~F. Dialektopoulos, Purba Mukherjee, Jackson Levi~Said, and Jurgen Mifsud.
\newblock {Neural network reconstruction of cosmology using the Pantheon compilation}.
\newblock {\em Eur. Phys. J. C}, 83(10):956, 2023.

\bibitem{Benisty:2022psx}
David Benisty, Jurgen Mifsud, Jackson Levi~Said, and Denitsa Staicova.
\newblock {On the robustness of the constancy of the Supernova absolute magnitude: Non-parametric reconstruction \& Bayesian approaches}.
\newblock {\em Phys. Dark Univ.}, 39:101160, 2023.

\bibitem{Chevallier:2000qy}
Michel Chevallier and David Polarski.
\newblock {Accelerating universes with scaling dark matter}.
\newblock {\em Int. J. Mod. Phys. D}, 10:213--224, 2001.

\bibitem{Linder:2005ne}
Eric~V. Linder and Dragan Huterer.
\newblock {How many dark energy parameters?}
\newblock {\em Phys. Rev. D}, 72:043509, 2005.

\bibitem{Tierney:1986}
Luke Tierney and Joseph~B. Kadane.
\newblock Accurate approximations for posterior moments and marginal densities.
\newblock {\em Journal of the American Statistical Association}, 81(393):82--86, 1986.

\bibitem{Lewis:2002ah}
Antony Lewis and Sarah Bridle.
\newblock {Cosmological parameters from CMB and other data: A Monte Carlo approach}.
\newblock {\em Phys. Rev. D}, 66:103511, 2002.

\bibitem{DESI:2024cru}
E.~Paillas et~al.
\newblock {Optimal reconstruction of baryon acoustic oscillations for DESI 2024}.
\newblock {\em JCAP}, 01:142, 2025.

\bibitem{Cowan:2010js}
Glen Cowan, Kyle Cranmer, Eilam Gross, and Ofer Vitells.
\newblock {Asymptotic formulae for likelihood-based tests of new physics}.
\newblock {\em Eur. Phys. J. C}, 71:1554, 2011.
\newblock [Erratum: Eur.Phys.J.C 73, 2501 (2013)].

\bibitem{Sellentin:2015waz}
Elena Sellentin and Alan~F. Heavens.
\newblock {Parameter inference with estimated covariance matrices}.
\newblock {\em Mon. Not. Roy. Astron. Soc.}, 456(1):L132--L136, 2016.

\bibitem{Tegmark:1996qt}
Max Tegmark.
\newblock {How to measure CMB power spectra without losing information}.
\newblock {\em Phys. Rev. D}, 55:5895--5907, 1997.

\bibitem{Heavens:2009nx}
Alan Heavens.
\newblock {Statistical techniques in cosmology}.
\newblock 6 2009.

\bibitem{Heavens:2007ka}
Alan~F. Heavens, T.~D. Kitching, and L.~Verde.
\newblock {On model selection forecasting, Dark Energy and modified gravity}.
\newblock {\em Mon. Not. Roy. Astron. Soc.}, 380:1029--1035, 2007.

\bibitem{Takahashi:2009ty}
Ryuichi Takahashi, Naoki Yoshida, Masahiro Takada, Takahiko Matsubara, Naoshi Sugiyama, Issha Kayo, Takahiro Nishimichi, Shun Saito, and Atsushi Taruya.
\newblock {Non-Gaussian Error Contribution to Likelihood Analysis of the Matter Power Spectrum}.
\newblock {\em Astrophys. J.}, 726:7, 2011.

\bibitem{Vallisneri:2011ts}
Michele Vallisneri.
\newblock {Beyond Fisher: exact sampling distributions of the maximum-likelihood estimator in gravitational-wave parameter estimation}.
\newblock {\em Phys. Rev. Lett.}, 107:191104, 2011.

\bibitem{Wolz:2012sr}
Laura Wolz, Martin Kilbinger, Jochen Weller, and Tommaso Giannantonio.
\newblock {On the Validity of Cosmological Fisher Matrix Forecasts}.
\newblock {\em JCAP}, 09:009, 2012.

\bibitem{Seehars:2014ora}
Sebastian Seehars, Adam Amara, Alexandre Refregier, Aseem Paranjape, and Jo\"el Akeret.
\newblock {Information Gains from Cosmic Microwave Background Experiments}.
\newblock {\em Phys. Rev. D}, 90(2):023533, 2014.

\bibitem{Repp:2015jja}
Andrew Repp, Istv\'an Szapudi, Julien Carron, and Melody Wolk.
\newblock {The Impact of Non-Gaussianity upon Cosmological Forecasts}.
\newblock {\em Mon. Not. Roy. Astron. Soc.}, 454(4):3533--3541, 2015.

\bibitem{Foroozan:2021zzu}
Setareh Foroozan, Alex Krolewski, and Will~J. Percival.
\newblock {Testing large-scale structure measurements against Fisher matrix predictions}.
\newblock {\em JCAP}, 10:044, 2021.

\bibitem{Dumerchat:2022uxh}
Tyann Dumerchat and Julian~E. Bautista.
\newblock {Baryon acoustic oscillations from a joint analysis of the large-scale clustering in Fourier and configuration space}.
\newblock {\em Astron. Astrophys.}, 667:A80, 2022.

\bibitem{Ding:2023ibo}
Zhejie Ding, Yu~Yu, and Pengjie Zhang.
\newblock {Fisher forecast for the BAO measurements from the CSST spectroscopic and photometric galaxy clustering}.
\newblock {\em Mon. Not. Roy. Astron. Soc.}, 527:3728--3740, 2023.

\bibitem{Yahia-Cherif:2020knp}
S.~Yahia-Cherif, A.~Blanchard, S.~Camera, S.~Casas, S.~Ili\'c, K.~Markovic, A.~Pourtsidou, Z.~Sakr, D.~Sapone, and I.~Tutusaus.
\newblock {Validating the Fisher approach for stage IV spectroscopic surveys}.
\newblock {\em Astron. Astrophys.}, 649:A52, 2021.

\bibitem{Note1}
\protect \url {https://github.com/dfm/emcee}.

\bibitem{Handley:2015fda}
W.~J. Handley, M.~P. Hobson, and A.~N. Lasenby.
\newblock {PolyChord: nested sampling for cosmology}.
\newblock {\em Mon. Not. Roy. Astron. Soc.}, 450(1):L61--L65, 2015.

\bibitem{Handley:2015vkr}
W.~J. Handley, M.~P. Hobson, and A.~N. Lasenby.
\newblock {polychord: next-generation nested sampling}.
\newblock {\em Mon. Not. Roy. Astron. Soc.}, 453(4):4385--4399, 2015.

\bibitem{Benisty:2024lmj}
David Benisty, Supriya Pan, Denitsa Staicova, Eleonora Di~Valentino, and Rafael~C. Nunes.
\newblock {Late-time constraints on interacting dark energy: Analysis independent of H0, rd, and MB}.
\newblock {\em Astron. Astrophys.}, 688:A156, 2024.

\end{thebibliography}
\end{document}